\documentclass[twocolumn]{aastex63}

\usepackage{epsfig}
\usepackage{amsmath}
\usepackage{amsfonts}
\usepackage{amssymb}
\usepackage{bm}
\usepackage{color}
\usepackage{graphicx}
\usepackage{subfigure}
\usepackage{ulem}

\newcommand{\G}{{\cal G}}
\renewcommand{\r}{{\rho}}

\newcommand{\s}{{\sum}}

\newcommand{\f}{{f^{(0)}}}

\newcommand{\p}[1]{\partial_{#1}}

\renewcommand{\vec}[1]{\bm {#1}}

\newcommand{\tij}[1]{\tau_{ij}^{\ell,#1}}
\newcommand{\ov}{\overline{v}}
\newcommand{\ob}{\overline{b}}

\newcommand{\ovpi}{\overline{v_i^+}}
\newcommand{\obpi}{\overline{b_i^+}}
\newcommand{\ovmi}{\overline{v_i^-}}
\newcommand{\obmi}{\overline{b_i^-}}

\newcommand{\ovsai}{\overline{v_i^{s_1}}}
\newcommand{\obsai}{\overline{b_i^{s_1}}}
\newcommand{\ovsbi}{\overline{v_i^{s_2}}}
\newcommand{\obsbi}{\overline{b_i^{s_2}}}
\newcommand{\ovscj}{\overline{v_j^{s_3}}}
\newcommand{\obscj}{\overline{b_j^{s_3}}}

\newcommand{\eps}{\varepsilon}

\newcommand{\be}{\begin{equation}}
\newcommand{\ee}{\end{equation}}
\newcommand{\bea}{\begin{eqnarray}}
\newcommand{\eea}{\end{eqnarray}}

\begin{document}

\title{Effects of forcing mechanisms on the multiscale properties of magnetohydrodynamics}

\author{Yan Yang}
\affiliation{Department of Mechanics and Aerospace Engineering, Southern University of Science and Technology, Shenzhen 518055, China}

\author{Moritz Linkmann}
\affiliation{School of Mathematics and Maxwell Institute for Mathematical Sciences, University of Edinburgh, Edinburgh EH9 3FD, United Kingdom}

\author{Luca Biferale}
\affiliation{Department of Physics and INFN, University of Rome ``Tor Vergata", Via della Ricerca Scientifica 1,
I-00133 Rome, Italy}

\author{Minping Wan}
\email{wanmp@sustech.edu.cn}
\affiliation{Department of Mechanics and Aerospace Engineering, Southern University of Science and Technology, Shenzhen 518055, China}
\affiliation{Guangdong Provincial Key Laboratory of Turbulence Research and Applications, Southern University of Science and Technology, Shenzhen, Guangdong 518055, China}



\begin{abstract}
We performed numerical simulations to study the response of magnetohydrodynamics (MHD) 
to large-scale stochastic forcing mechanisms parametrized by one parameter, $0 \le a \le1$, 
going from direct injection on the velocity field ($a = 1$) to 
stirring acts on the magnetic field only ($a = 0$).
We study the multi-scale properties of the energy transfer, by splitting the total flux in channels mediated by 
(i) the kinetic non-linear advection, (ii) the Lorentz force, (iii) the magnetic advection and (iv) magnetic stretching term. 
We further decompose the fluxes in two sub-channels given by heterochiral and homochiral components in order to distinguish 
forward, inverse and flux-loop cascades. 
We show that there exists a quasi-singular role of the magnetic forcing mechanism for $a \sim 1$:
a small injection on the magnetic field $a < 1$ can strongly deplete the mean flux of 
kinetic energy transfer throughout the kinetic non-linear advection channel. 
We also show that this negligible mean flux is the result of a flux-loop balance 
between heterochiral (direct) and homochiral (inverse) transfers. 
Conversely, both homochiral and heterochiral channels transfer energy forward for the other three channels. 
Cross exchange between velocity and the magnetic field is reversed around $a = 0.4$ and 
except when $a \sim 1$ we always observe that heterochiral mixed velocity-magnetic energy triads tend to move energy from magnetic to velocity fields. 
Our study is an attempt to further characterize the multi-scale nature of MHD dynamics, 
by disentangling different properties of the total energy transfer mechanisms, which can be useful for improving sub-grid-modelling.
\end{abstract}



\section{Introduction} \label{sec:intro}

The turbulent dynamics of conducting fluids such as liquid metals and plasmas
in the one-fluid approximation, are relevant to a variety of observed phenomena
in astro- and geophysical fluid dynamics.  The most pertinent examples are
stellar and planetary dynamos \citep{Moffatt78,Parker79,Brandenburg05,Tobias13}, the granulation in the solar convection zone \citep{leighton1962velocity,leighton1963solar,bray1984solar,rieutord2010sun}, and turbulence in the solar wind
and Earth's magnetosheath \citep{Barnes79a,tu1995mhd,goldstein1995magnetohydrodynamic,bruno2005solar}.  All these systems can be described to a good
approximation by the equations of magnetohydrodynamics (MHD), albeit at
different parameter ranges, and potentially coupled to additional evolution
equations.  The magnetic activity of the Sun, for instance, originates from
the solar tachocline \citep{miesch2005large} and the main features
of its dynamics can be well described by an ideal, i.e. dissipationless, MHD
shallow-water model \citep{gilman2000magnetohydrodynamic,dikpati2001prolateness}.  The latter highlights an important unresolved challenge
in astrophysical fluid dynamics - both magnetic and fluid Reynolds numbers are
usually very high, leading to turbulent dynamics on scales much too small to be
adequately resolved in numerical simulations. The parameter ranges are way
beyond the capabilites even for optimised codes running on state-of-the-art
high-performance computing facilities. Observational studies give very
important insights \citep{tu1995mhd,horbury2005spacecraft,chen2016recent,bruno2013solar,matthaeus2019plasma,klein2019plasma}, however the data
acquisition process is complex and high-order statistics are 
difficult to measure precisely.  

To further disentangle the complexity of MHD turbulence, a better understanding
of the intricate multi-scale transfer of inertial conserved quantities as the
total energy, the magnetic and the cross helicity is mandatory
\citep{Mininni11,Biskamp03,galtier2016introduction,zhou2004colloquium,Alexakis2018cascades}.
The overall global picture might strongly depend on the injection properties,
being the three conserved quantities dynamically correlated. As a result,
systematic studies at varying one or a few control parameters, connected to the
properties of the stirring mechanisms are important for both fundamental and
applied interests, e.g. concerning the issue of universality - independence of
the forcing -  and small-scale modelling
\citep{alexakis2013large,bian2019decoupled,McKay17,Zhou91,muller2002,Chernyshov10,Kessar16,miesch2015}.

In recent years, it has become more and more clear that global measurements
based on spectra and mean fluxes do not allow for a precise disentanglement of
the physics mechanisms underlying the multi-scale transport, e.g.  direct or
inverse energy cascades can have the same spectral properties.
In MHD turbulence the situation is particularly complicated, as the interaction of magnetic and velocity fluctuations can 
proceed in different parameter regimes. A background magnetic field or different levels of 
cross- and magnetic helicities, for instance, have measureable effects on the value of the scaling exponent of  
magnetic and kinetic energy spectra. The complexity of the problem is reflected in the 
considerable effort that has been put upon measuring and understanding spectral scaling   
and other second-order statistics for the incompressible case
\citep{Iroshnikov64,Kraichnan65a,Matthaeus89,Goldreich95,Mueller00,galtier2000wave,galtier2002anisotropy,Boldyrev05a,Boldyrev06,mason2006dynamicalignment,Boldyrev09,Beresnyak09,Grappin10}. 
More recently, effects specific to compressible flows have been quantified \citep{teissier2020scaling,grete2020kinetic}.

Similarly, even
the sign of the total flux does not fully summarise the entire flow dynamics,
we know examples where the mean flux is vanishing because of the result of
strong counter-reacting positive and negative transfers, the so-called
flux-loop case, which is obviously very far from a quasi-equilibrium state
\citep{Alexakis2018cascades}. Similarly, in the purely hydrodynamic case of
Navier-Stokes equations, it is known that homochiral and heterochiral Fourier
interactions transfer energy in opposite directions across scales, even in
purely 3d homogeneous and isotropic turbulence \citep{Waleffe93,Biferale12}.
Such subtle refinements in the description of the energy transfer properties
are interesting also from an applied point of view, opening the way for
controlling both small and large scale flow behaviours by suitable forcing
properties, aimed to switch on/off some energy transfer sub-channel. 

In this paper we focus on the multi-scale properties of the energy transfer for 3d MHD under isotropic and homogeneous conditions. In particular, we present the results from  a series of direct numerical simulations to study  the response of the conducting fluid to a family of large-scale stochastic
	forcing mechanisms parametrised by one single parameter, $0 \le a \le
	1$, where $a=1$ means forcing only on the kinetic channel and  $a=0$ only on the magnetic one. We study the multi-scale
	properties of the energy transfer, using two different levels of splitting. First, we analyze the properties of the 
	four sub-classes (channels) given by  (i) the kinetic  non-linear
	advection, (ii) the Lorentz force, (iii) the magnetic advection and
	(iv) magnetic stretching term. Second, we further decompose each of the four fluxes in two sub-channels given by  heterochiral and homochiral components in order 	to distinguish forward, inverse and flux-loop cascades. 
The main results are: (i)  the mean energy transferred by the kinetic advection channel is strongly depleted as soon as  we switch on  a small injection on the magnetic field; (ii) this  negligible  mean  flux  is  the  result  of  a  flux-loop  balance between heterochiral (direct transfer) and homochiral (inverse transfer); (iii)   heterochiral transfers are strongly affected by the relative amount of magnetic/kinetic forcing; (iv)   we observed that the velocity-magnetic energy exchange mediated by  mixed triads tends to move energy from magnetic to velocity fields.\\
The article is organised as follows. In Sec. \ref{sec:numerics} we present our numerical dataset. In Sec. \ref{sec:flux_dec} we describe the results obtained by splitting  the total flux in four components and in Sec. \ref{sec:homo-hetero} how the ones obtained after a further decomposition   in heterochiral and homochiral channels. Results concerning the exchange between kinetic and magnetic fields can be found in Sec. \ref{sec:exchange} and the ones for small-scale fluctuations in Sec. \ref{sec:small_scale}

\section{Description of the dataset}
\label{sec:numerics}
\begin{figure}[h]
	\centering
	 \vspace{0.5em}
	\includegraphics[width = \columnwidth]{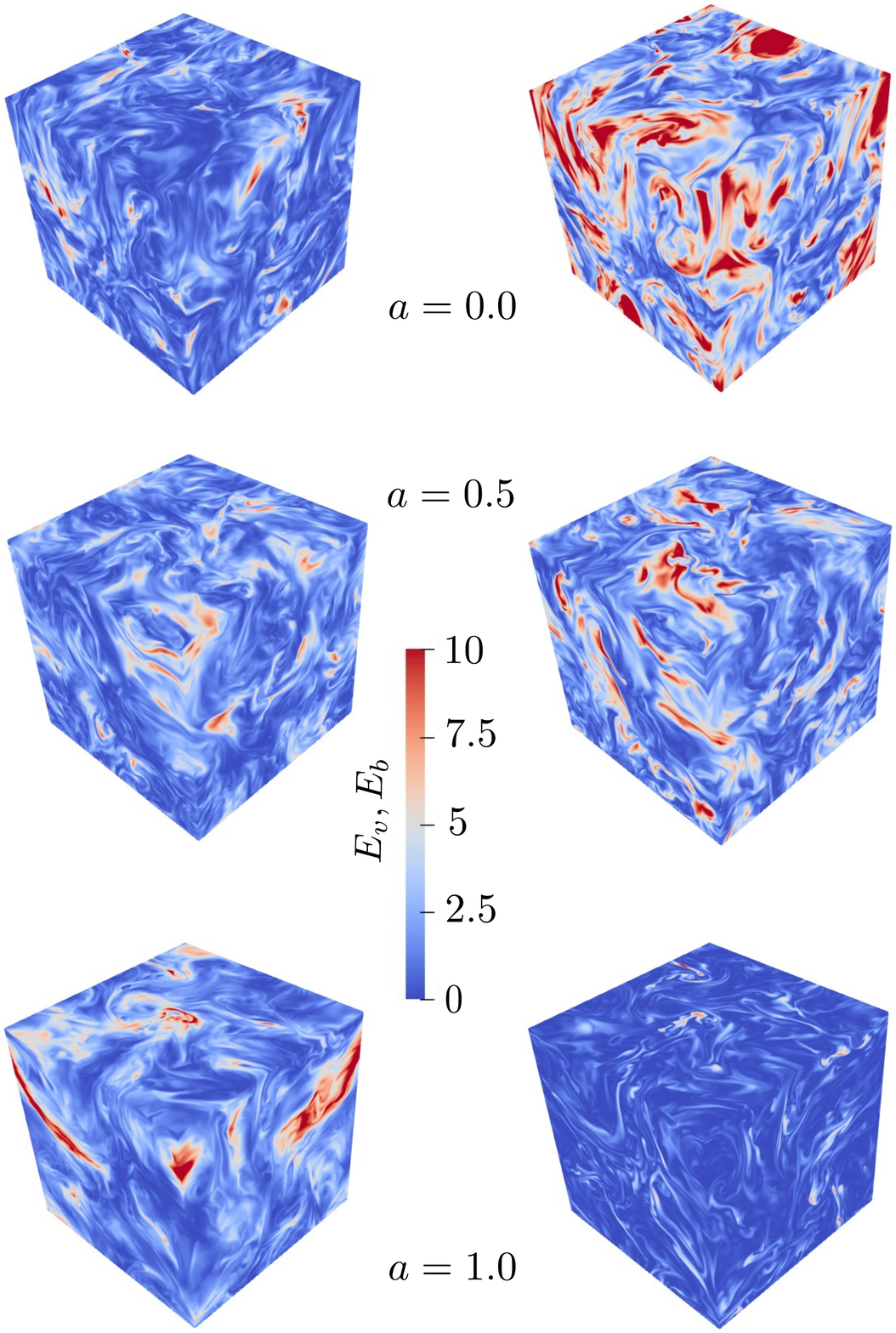}
	\caption{Visualizations of (unaveraged) kinetic (left)
	and (unaveraged) magnetic (right) energy 
	 distributions for $a = 0.00$, $a = 0.50$ and $a = 1.00$ at $t/T = 6.9$. }
	\label{fig:3drendering}
\end{figure}

To study the effect of large-scale forcing mechanisms on the multiscale properties of MHD,
we conduct direct numerical simulations of the incompressible MHD equations 
\begin{align}
	\label{eq:momentum}
\frac{\partial \bm{v}}{\partial t} + \left(\bm{v} \cdot \nabla\right) \bm{v} &=-\nabla p_M + \left(\bm{b} \cdot \nabla\right) \bm{b} + \nu \Delta \bm{v} + \sqrt{a} \bm{f}_v,\\
	\label{eq:induction}
\frac{\partial \bm{b}}{\partial t} + \left(\bm{v} \cdot \nabla\right) \bm{b} &= \left(\bm{b} \cdot \nabla\right) \bm{v} + \eta \Delta \bm{b} + \sqrt{1-a} \bm{f}_b, \\
	\label{eq:incomp}
	\nabla \cdot \bm{v} = 0, & \quad  \nabla \cdot \bm{b}=0.
\end{align}
where $\bm{v}$, $\bm{b}$, $p_M=p+|\bm{b}|^2/2$, $\nu$ and $\eta$
denote the velocity, the magnetic field, the total pressure, the kinematic viscosity
and the magnetic resistivity, respectively.
The velocity and magnetic fields are driven respectively by $\sqrt{a} \bm{f}_v$ and $\sqrt{1-a} \bm{f}_b$ at the first two wave numbers,
where $\bm{f}_v$ and $\bm{f}_b$ are random, Gaussian-distributed and $\delta(t)$-correlated forces and
$\langle |\bm{f}_v|^2\rangle = \langle |\bm{f}_b|^2\rangle$.
As a result the total energy input:  
\be
  a \langle |\bm{f}_v|^2\rangle + (1-a) \langle |\bm{f}_b|^2\rangle = \epsilon^{in}_v + \epsilon_{b}^{in} = \epsilon^{in} =  const. \
\ee
 is independent of $a$.
Modifying the parameter $a$ allows us to construct an injection mechanism  from purely mechanical, $a=1$, to purely magnetic, $a=0$. A first visual understanding of our set-up can be seen in Fig. \ref{fig:3drendering}, where we show a 3d rendering of kinetic and magnetic energy distribution in the whole volume for three characteristic values, $a=0,0.5,1$. The presence of some unbalance between purely kinetic and purely magnetic forcing is detectable even by naked eyes, our goal is to quantify and go deeper into the entangled dynamics at changing the forcing properties $a$.

\begin{figure}
    \includegraphics[width = .98\columnwidth]{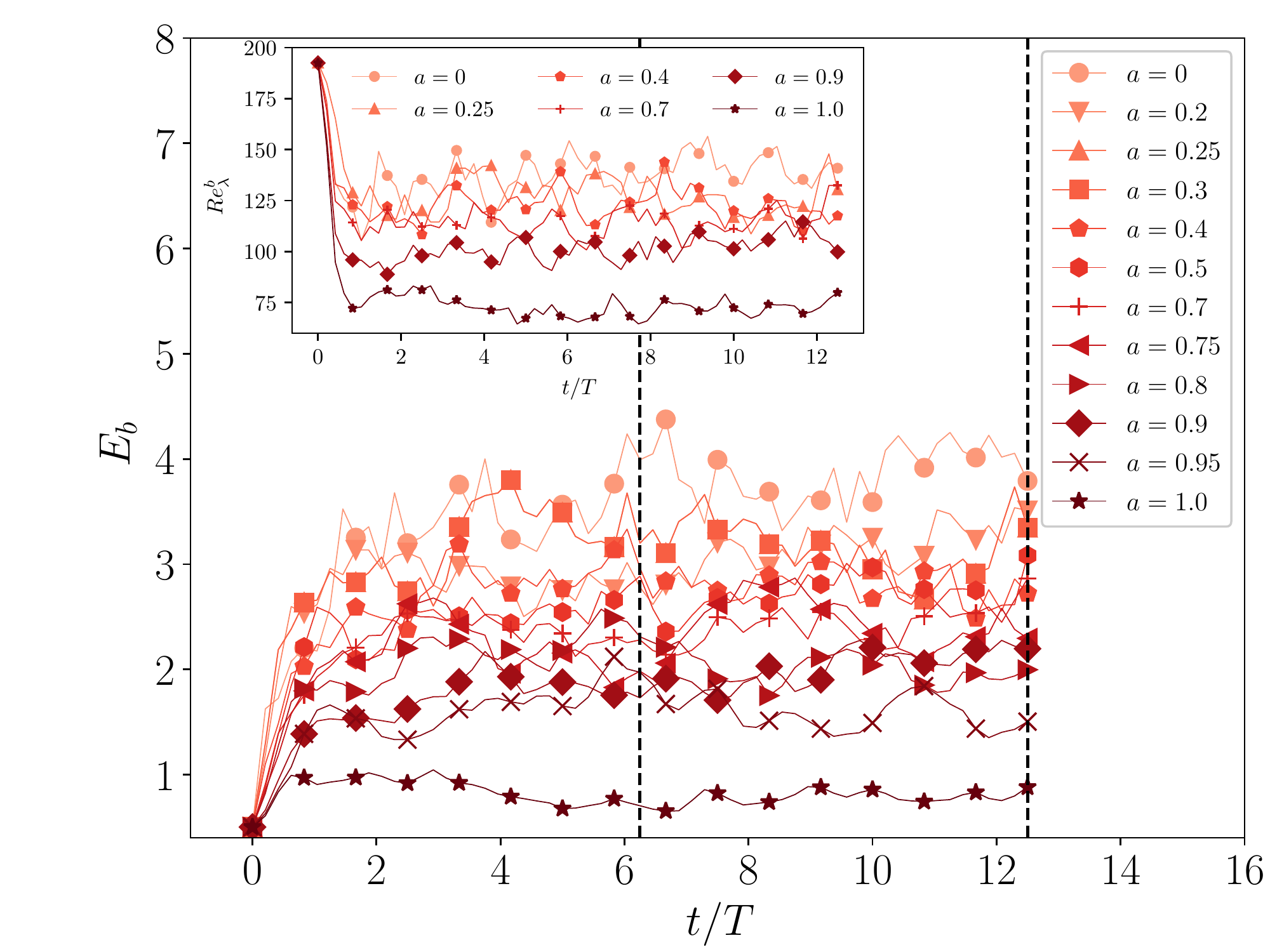}
	\includegraphics[width = .98\columnwidth]{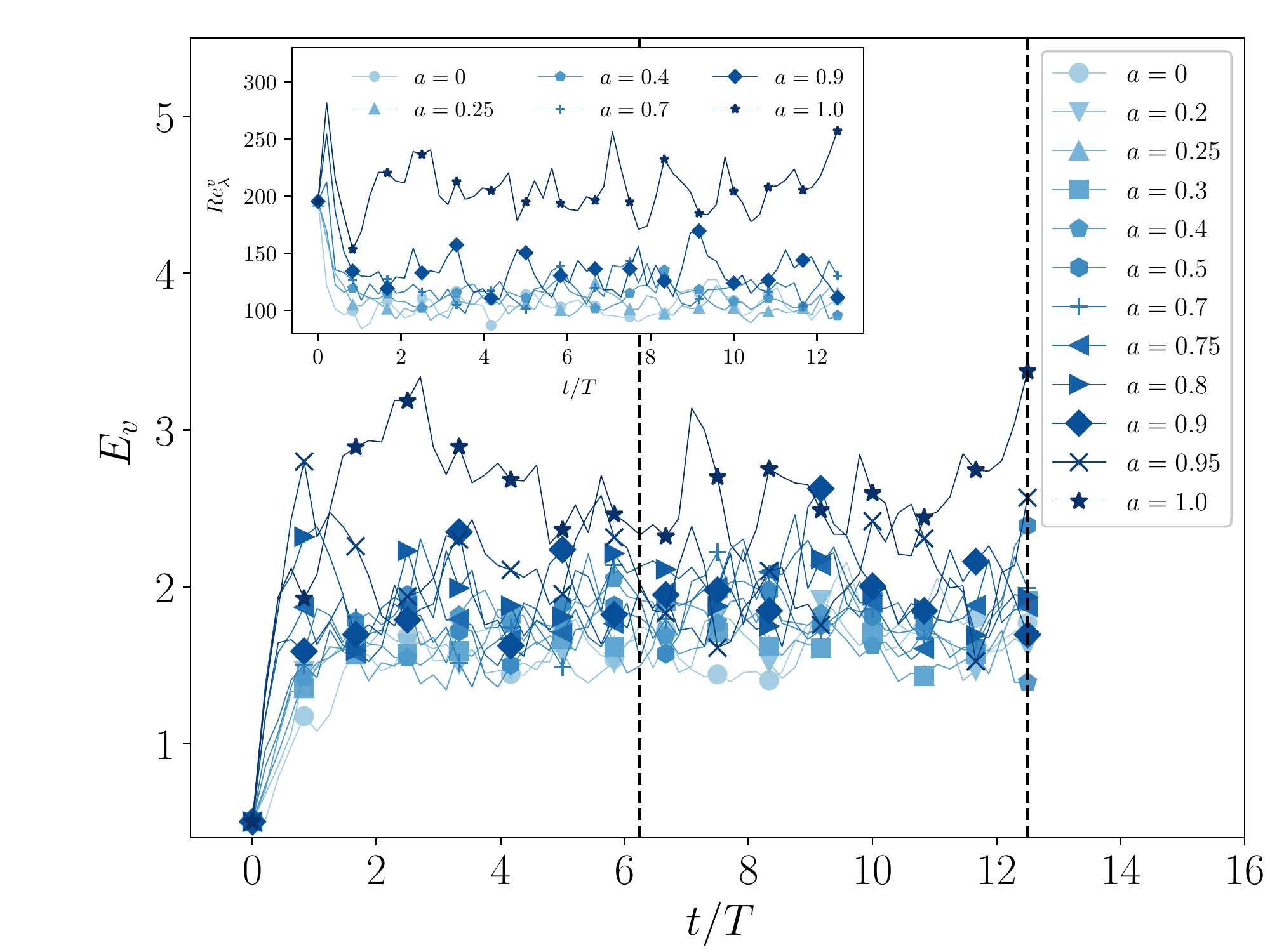}
	\caption{Time evolution of magnetic (top) and kinetic (bottom) energies as functions of $a$. 
	Insets: time evolution of the Taylor-scale Reynolds numbers as functions of $a$ are shown in the insets. The interval used in the data analyses is indicated by the vertical lines.}
	\label{fig:energy-timeseries}
\end{figure}

From a practical point of view, we numerically solved 
eqs.~\eqref{eq:momentum}-\eqref{eq:incomp} 
on a cubic domain $\Omega = [0,2\pi]^3$ with periodic boundary conditions
in all directions using
the standard pseudo-spectral method with dealiasing by the two-thirds rule. The fields were
advanced in time by a second-order Adam-Bashforth scheme.
All runs discussed here were carried out using $512$ collocation points in each direction.
The simulations were initialized by supplying randomly generated velocity and magnetic field fluctuations
at wavenumbers $1\le |\bm{k}| \le 5$,
with spectra proportional to
$1/\left[1+(k/k_0)^{11/3}\right]$ with $k_0=3$. 
Magnetic and kinetic energies were initially in equipartition in all cases, where $E_v = 0.5$ and $E_b = 0.5$. 
The magnetic Prandtl number is $\text{Pm} = \nu/\eta =1$ for all simulations, and we did not impose an external mean  magnetic field.
The cross-helicity $\sigma_c=2\langle \bm{v} \cdot \bm{b}\rangle/\langle |\bm{v}|^2 + |\bm{b}|^2\rangle$,
though not exactly zero, is very small.
So here we will be studying the energy transfer in simplified cases,
lacking complexities that cross-helicity might introduce.
In Fig.~\ref{fig:energy-timeseries} we show the time evolution of the
kinetic and magnetic energy at changing the control parameter, $a$. As one can
see, all runs reached a stationary state after a time of the order of $\sim 2
T$, where $T=L_v/U$ is the large-eddy turnover time, $U=\sqrt{2E_v/3}$ is the
r.m.s. velocity, $L_v=\left[\sum_k k^{-1} E_v(k)\right]/E_v$ is the kinetic
integral length scale and $E_v(k)$ is the kinetic energy spectrum. Note that
the large-eddy turnover times for the runs are different (see Table
\ref{tab:params}), and  the time throughout the paper will be in units of the
characteristic time $T$ of R1. All our  analysis will be limited in the
statistically stationary interval, i.e. from $t/T=6.25$ to $t/T=12.5$, where we used 31
equispaced snapshots. Further details on simulation parameters and main
observables are summarised in Table \ref{tab:params}.

\begin{table*}[h]
    \centering
    \begin{tabular}{ccccccccccccc}
    \hline
	    id  & $N$ & $a$  & $Re_L^{v}$ & $Re_L^{b}$ & $Re_{\lambda}^{v}$ & $Re_{\lambda}^{b}$ & $E_v$ & $E_b$ & $\eps_v$ & $\eps_b$ & $T$ & $k_{\rm max}\eta_v$ \\
     \hline
	    R1    & 512 & 0.00 & 1663 & 4506 & 105 & 141 & 1.69  & 3.87 & 0.87 & 1.10 & 2.40 & 1.65 \\
	    R2    & 512 & 0.20 & 1665 & 3366 & 105 & 127 & 1.64  & 3.14 & 0.82 & 1.07 & 2.47 & 1.68 \\
	    R3    & 512 & 0.25 & 1675 & 3317 & 105 & 126 & 1.67  & 3.16 & 0.84 & 1.10 & 2.44 & 1.67 \\
	    R4    & 512 & 0.30 & 1694 & 3224 & 106 & 128 & 1.63  & 2.92 & 0.79 & 1.03 & 2.54 & 1.70 \\
	    R5    & 512 & 0.40 & 1802 & 2865 & 109 & 122 & 1.69  & 2.81 & 0.80 & 1.07 & 2.59 & 1.69 \\
	    R6    & 512 & 0.50 & 2081 & 2721 & 117 & 122 & 1.83  & 2.72 & 0.82 & 1.12 & 2.75 & 1.68 \\
	    R7    & 512 & 0.70 & 2244 & 2247 & 122 & 115 & 1.88  & 2.42 & 0.80 & 1.16 & 2.90 & 1.69 \\
	    R8    & 512 & 0.75 & 2376 & 2269 & 125 & 116 & 1.94  & 2.43 & 0.80 & 1.17 & 2.99 & 1.69 \\
	    R9    & 512 & 0.80 & 2585 & 1731 & 131 & 107 & 1.96  & 2.02 & 0.75 & 1.16 & 3.21 & 1.72 \\
	    R10   & 512 & 0.90 & 2716 & 1630 & 134 & 104 & 2.00  & 2.01 & 0.75 & 1.24 &  3.29 & 1.72 \\
	    R11   & 512 & 0.95 & 3148 & 1162 & 144 & 94  & 2.04  & 1.60 & 0.68 & 1.24 & 3.72 & 1.76 \\
	    R12   & 512 & 1.00 & 6486 & 427  & 207 & 72  & 2.60  & 0.78 & 0.53 & 1.30 & 6.07 & 1.87 \\
     \hline
    \end{tabular}
    \caption{Parameters and key observable for all simulations: grid size $N^3$, forcing parameter $a$, kinetic and magnetic
    integral-scale Reynolds numbers $Re_L^{v}=U L_v /\nu$ and $Re_L^{b}=U L_b /\eta$ where $U=\sqrt{2E_v/3}$ is the rms velocity and $L_v=\left[\sum_k k^{-1} E_v(k)\right]/E_v$ and $L_b=\left[\sum_k k^{-1} E_b(k) \right]/E_b$ are the kinetic and magnetic integral length scales respectively, kinetic and magnetic Taylor-scale Reynolds numbers $Re_{\lambda}^{v}=U \lambda_v /\nu$ and $Re_{\lambda}^{b}=U \lambda_b /\eta$ where $\lambda_v$ and $\lambda_b$ are the kinetic and magnetic Taylor microscales respectively, kinetic energy $E_v$, magnetic energy $E_b$, kinetic dissipation rate $\eps_v$, magnetic dissipation rate $\eps_b$, large-eddy turnover time $T=L_v/U$, the largest resolved wavenumber $k_{\rm max}=N/3$, and kinetic Kolmogorov microscale $\eta_v=(\nu^3/\eps_v)^{1/4}$. All observables are time averaged.}
    \label{tab:params}
\end{table*}

\begin{figure}[h]
    \includegraphics[width = .98\columnwidth]{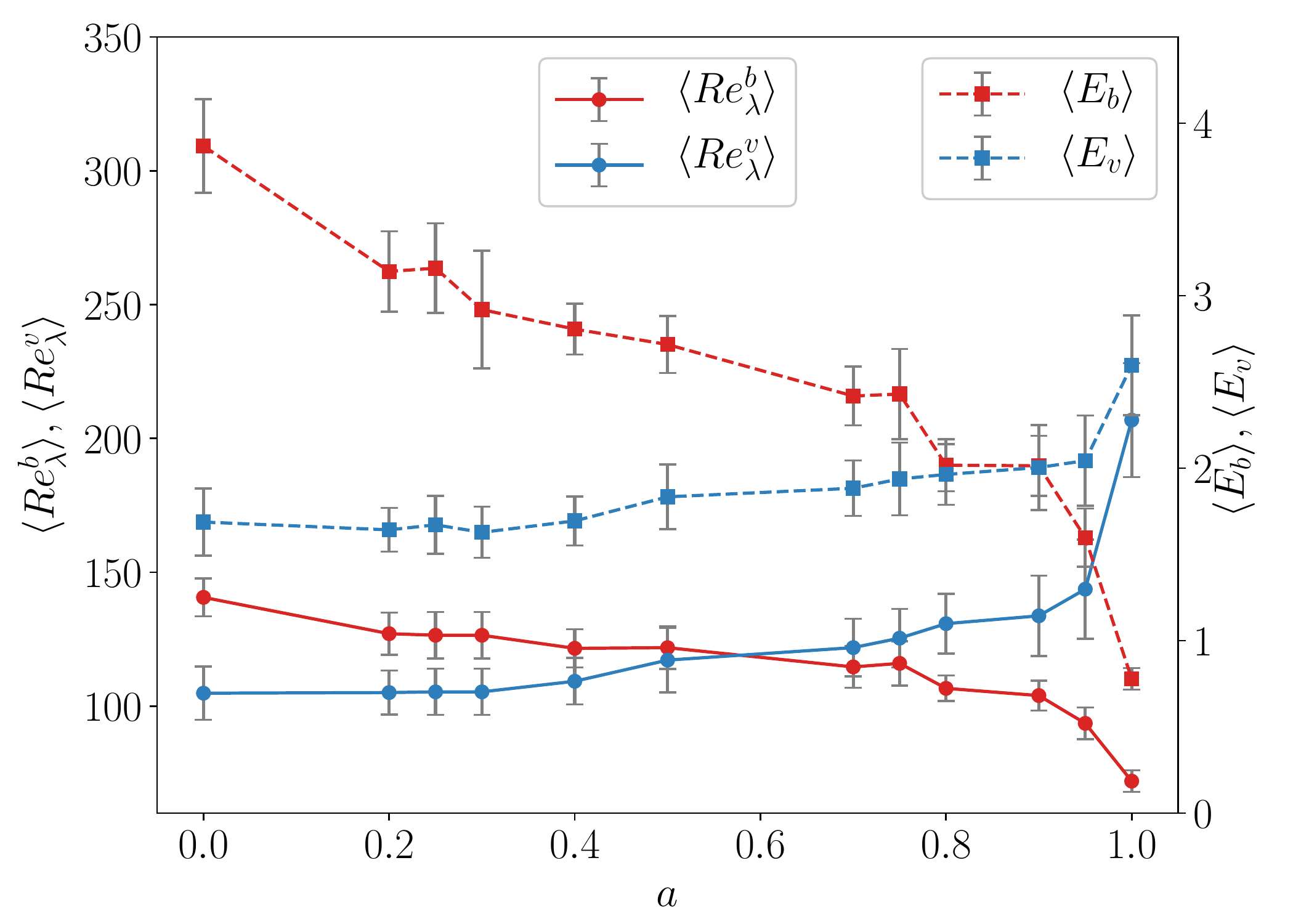}
	\caption{Time-averaged Taylor-scale Reynolds numbers and
	magnetic and kinetic energies as functions of $a$.
	The error bars show the standard error.}
	\label{fig:Re}
\end{figure}

\begin{figure}[h]
	\centering
    \includegraphics[width = .98\columnwidth]{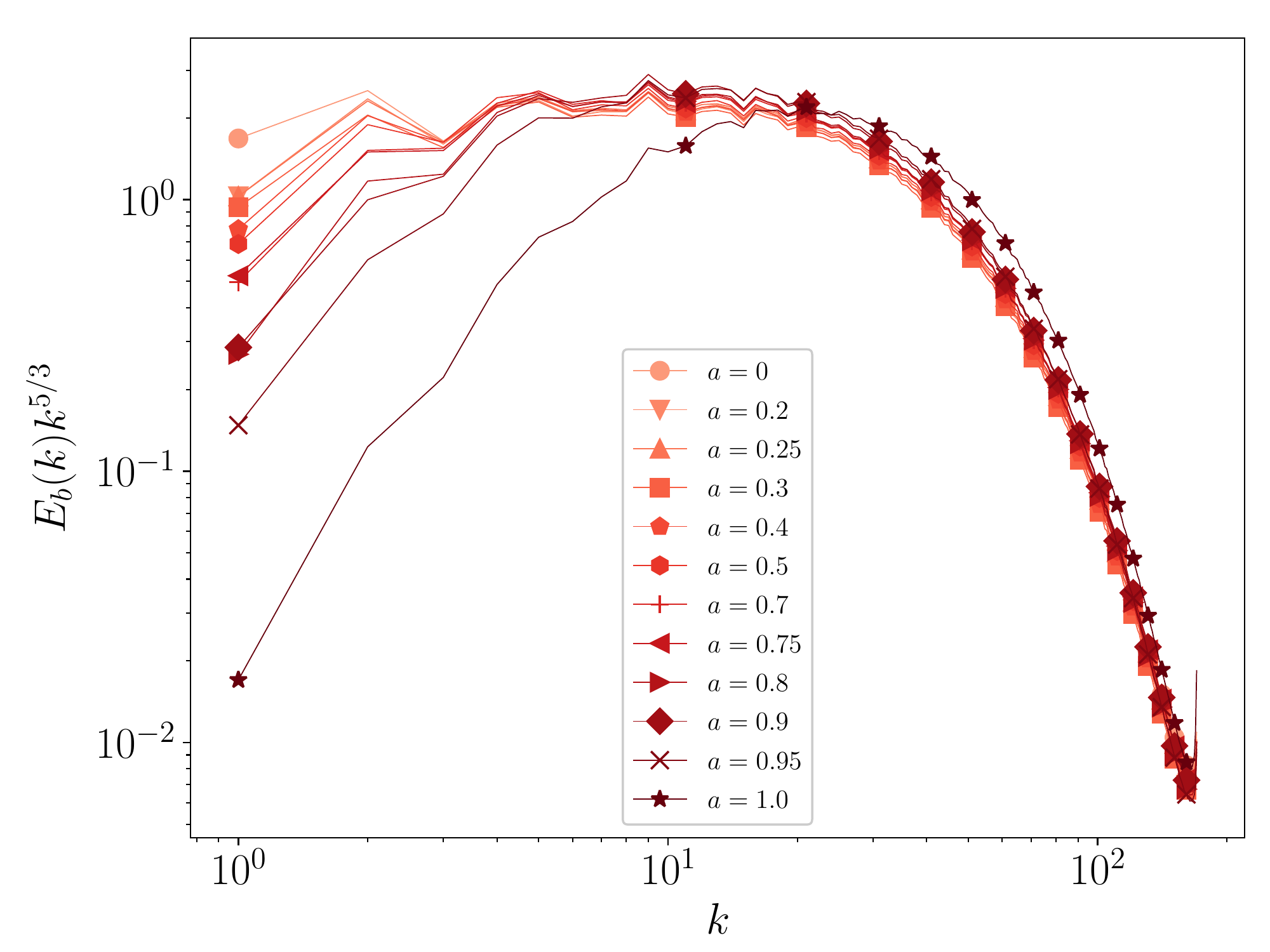}
    \includegraphics[width = .98\columnwidth]{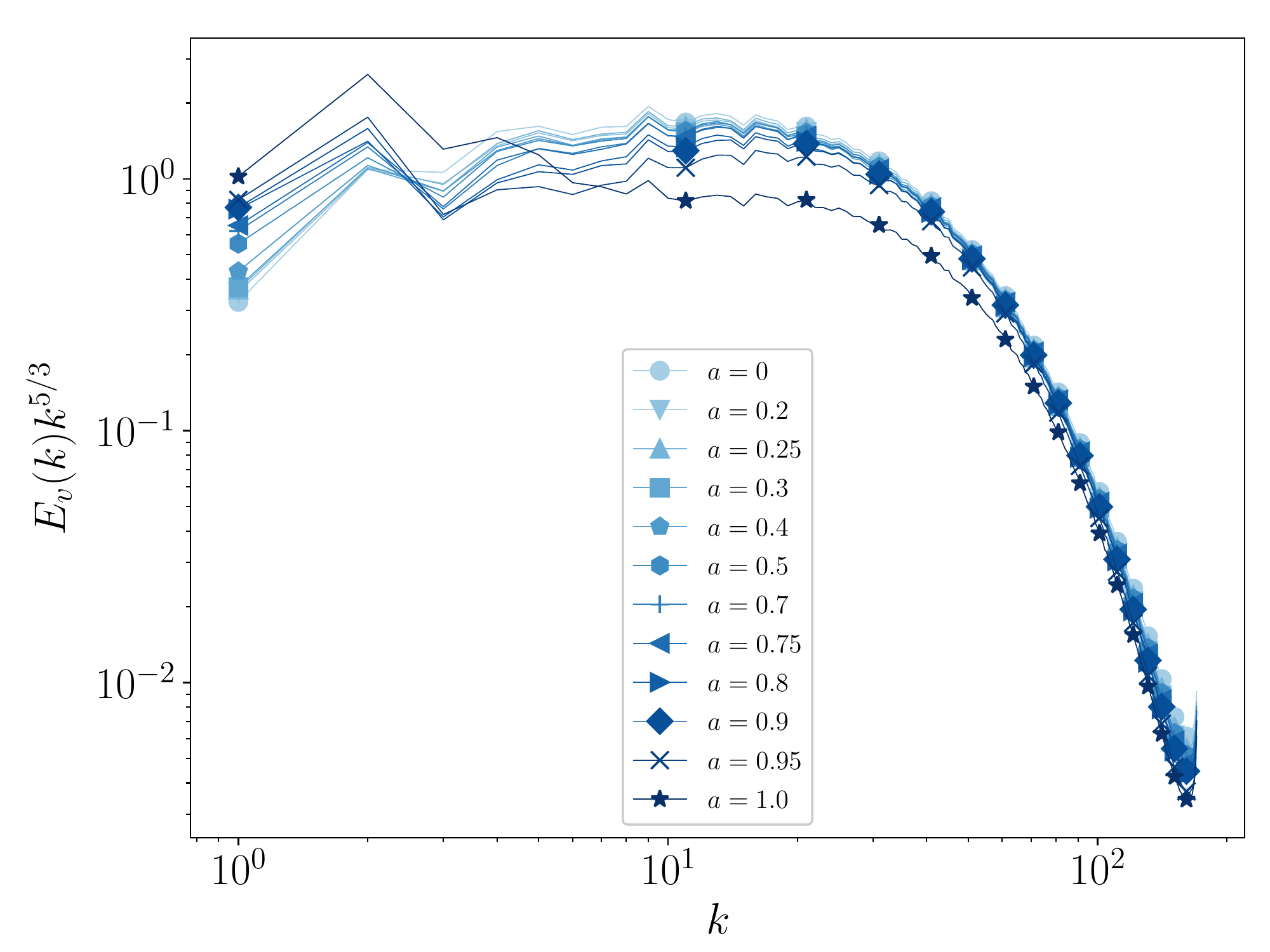}
	\caption{Magnetic (top) and kinetic (bottom) compensated energy spectra for different values of $a$, where $a = 1.00$ corresponds to fully mechanical forcing and $a = 0.00$ to fully electromagnetic forcing. The spectra have been averaged over 31 configurations during the statistically stationary evolution.}
	\label{fig:energy_spectra}
\end{figure}

In the inset of Fig. \ref{fig:energy-timeseries} we show also the evolution of the kinetic and magnetic Taylor-scale Reynolds numbers, 
\begin{align}
    Re_{\lambda}^{v} = \frac{U \lambda_v} {\nu}, \\
    Re_{\lambda}^{b} = \frac{U \lambda_b} {\eta}, 
\end{align}
where $\lambda_v=U/\omega_{\rm rms}$ and $\lambda_b=B/j_{\rm rms}$ are the kinetic and magnetic Taylor microscales respectively,
$B=\sqrt{2E_b/3}$ is the r.m.s. magnetic magnitude, and $\omega_{\rm rms}$ and  $j_{\rm rms}$ are the rms vorticity $\omega=|\nabla \times \bm{v}|$ and current density $j=|\nabla \times \bm{b}|$ respectively. In Fig. \ref{fig:Re} one can find the averaged values for all the above quantities, evaluated on the stationary regime as indicated by the vertical dashed  lines in Fig. \ref{fig:energy-timeseries}.
Fig. \ref{fig:energy_spectra} shows the kinetic and magnetic energy spectra 
\begin{align}
    E_v(k) = \frac{1}{2}  \sum_{|\bm{k}|=k} \left \langle |\hat{\bm{v}}(\bm{k},t)|^2\right \rangle_t  , \\
    E_b(k) = \frac{1}{2}  \sum_{|\bm{k}|=k} \left \langle |\hat{\bm{b}}(\bm{k},t)|^2 \right \rangle_t,
\end{align}
where $\langle \cdots \rangle_t$ denotes a time average over 31 snapshots and $\hat{\cdot}$ the Fourier transform.  

As can be seen from the spectra, the variation of $a$ is mostly visible at the large magnetic scales, while the 
large-scale dynamics of the velocity field is weakly affected, at least at the level of energy spectra. The kinetic energy
spectrum shows some difference in the inertial-range scaling and amplitude. For both magnetic and kinetic energy spectra, we 
note that $a=1.00$, that is, fully mechanical forcing is quasi-singular in appearance, in the sense that significant differences
in the spectra are discernible between $a=1.00$ and $a=0.95$. For smaller values of $a$ the spectra show little variation. The overall picture emerging from these few quantitative results confirm what already said in the previous text: it is very difficult, if not impossible, to disentangle the intricate multi-scale properties of the conducting flow by looking only at mean global properties based on spectra. In order to overcome this limitation, in the following section we will introduce the multi-channel flux decomposition as anticipated in the introduction.

\section{Flux decomposition: kinetic, Lorentz, magnetic advection, magnetic stretching}
\label{sec:flux_dec}
In order to study the multi-scale energy dynamics all vector fields must be decomposed into 
large and small-scale components. To do so, we apply a filtering operation to all terms in 
eqs.~\eqref{eq:momentum} and \eqref{eq:induction} \citep{Zhou91,Kessar16,Yang16a,Aluie17,Offermans18}. In what follows, we briefly summarise the main procedure, 
for the derivations and further details see \citep{Kessar16,Aluie17,Offermans18}.
Given a filter kernel $G^\ell$, the filtered component of a square-integrable function $\varphi(\vec{x})$ 
in the domain $\Omega$, is defined as
\begin{align}
  \overline{\varphi}^\ell(\vec{x}) \equiv
        \int_{\Omega}d\vec{y} \ G^\ell\left(\vec{x} - \vec{y}\right) \varphi(\vec{y}) =
  \sum\limits_{\vec{k} \in \mathbb{Z}^3} \hat{G}^\ell(\vec{k}) \hat{\varphi} (\vec{k}) e^{i\vec{k}\vec{x}} , \
  \label{eq:filtered_function}
\end{align}
where  $\ell$ is the filter width. Several choices of $G$, such as Gaussian filters, top hat filters or Galerkin projectors may be used to decompose a given 
function into large-scale and small-scale components. For the present study, $G^\ell$ is a spherically symmetric Galerkin projector
\be
\label{eq:filter}
\hat{G}^\ell(\vec{k}) = 
\begin{cases}
	1 & \text{if } |\vec{k}| \leqslant k_c \ , \\
	0 & \text{if } |\vec{k}| > k_c \ , 
\end{cases}
\ee
where $k_c = \pi/\ell$. Galerkin projectors have a number of advantages and disadvantages. On the positive side, the projection operation results in a clear distinction 
between sub-filter and resolved scales, and the sub-filter-scale energy fluxes calculated with respect to the projector coincide exactly with the classically defined Fourier 
fluxes. Drawbacks are mainly concerned with filter-induced oscillations in configuration space that result in sub-filter-scale stresses that are not positive definite \citep{Vreman94a}.
However, the statistics of the sub-filter-scale energy transfer calculated through Gaussian smoothing or Galerkin projection vary little for Navier-Stokes turbulence \citep{Buzzicotti18a}.

The filtered MHD equations read 
\begin{align}
	\p{t}\ov_i^\ell =& - \p{j}\left(\overline{\ov_i^\ell \ov_j^\ell}^\ell - \overline{\ob_i^\ell\ob_j^\ell}^\ell + \tij{I} - \tij{M} + \overline{p}^\ell\delta_{ij}\right) \nonumber\\
	                 & + \nu \p{jj} \ov_i^\ell + \sqrt{a} \, \overline{{f_v}_i}^\ell \ , \label{eq:momentum_filtered}\\
	\p{t}\ob_i^\ell =& - \p{j}\left(\overline{\ob_i^\ell \ov_j^\ell}^\ell  - \overline{\ov_i^\ell \ob_j^\ell }^\ell  + \tij{A}  - \tij{D}  \right) \nonumber\\
	                 & + \eta \p{jj} \ob_i^\ell  +  \sqrt{1-a} \, \overline{{f_b}_i}^\ell \ , \label{eq:induction_filtered}
\end{align}
where we sum over repeated indices and 
\begin{align}
        \tij{I} =& \overline{v_i v_j}^\ell - \overline{\ov_i^\ell \ov_j^\ell}^\ell \ , \label{eq:SGS_tensor_I} \\
	\tij{M} =& \overline{b_i b_j}^\ell - \overline{\ob_i^\ell \ob_j^\ell}^\ell \ , \label{eq:SGS_tensor_M} \\
	\tij{A} =& \overline{b_i v_j}^\ell - \overline{\ob_i^\ell \ov_j^\ell}^\ell \ , \label{eq:SGS_tensor_A} \\
	\tij{D} =& \overline{v_i b_j}^\ell - \overline{\ov_i^\ell \ob_j^\ell}^\ell \ , \label{eq:SGS_tensor_D}
\end{align}
denote the inertial (I), Maxwell (M), advective (A) and dynamo (D) subfilter-scale stresses, respectively. Despite their common origin through the electric field in the induction 
equation, we here treat $\tij{A}$ and $\tij{D}$ separately, in order to disentangle the effects of magnetic-field-line advection, encoded in $\tij{A}$, 
and magnetic-field-line stretching, encoded in $\tij{D}$. Usually, the magnetic sub-scale stress refers to the difference $\tij{A} - \tij{D}$ \citep{Aluie17,Offermans18}. 
Equations \eqref{eq:momentum_filtered} and \eqref{eq:induction_filtered} differ from expressions for the filtered MHD equations found elsewhere by an additional projection of the 
coupling terms. The latter ensures that the dynamics defined by eqs.~\eqref{eq:momentum_filtered} and \eqref{eq:induction_filtered} are confined to the same finite-dimensional 
subspace $\Omega^\ell$ of the original domain $\Omega$ \citep{Buzzicotti18a,Offermans18}. At first sight, this formulation suggests that the corresponding evolution equations 
for kinetic and magnetic energy feature terms that are not Galilean invariant, which ought to be avoided as the measured subfilter-scale energy transfers otherwise 
include unphysical fluctuations \citep{aluie2009I,aluie2009II,Buzzicotti18a}. However, the energy balance equations can be expressed in an alternative way by including terms that 
vanish under spatial averaging and ensure Galilean invariance of all terms \citep{Buzzicotti18a,Offermans18}. 
For a statistically stationary evolution, the spatio-temporally averaged energy budget can then be written as 
\begin{align}
  \label{eq:evol_Eu}
 \sqrt{a}	\left \langle \ov_i^\ell \overline{{f_v}_i}^\ell \right  \rangle &=   \Pi_I^\ell  -  \Pi_M^\ell  - \pi_M^\ell  + \eps_v^\ell\ , \\
  \label{eq:evol_Eb}
	\sqrt{1-a} \, \left \langle \ob_i^\ell \overline{{f_b}_i}^\ell  \right \rangle &=   \Pi_A^\ell  -  \Pi_D^\ell   -\pi_D^\ell  +  \eps_b^\ell\ , 
\end{align}
where $\eps_v^\ell = \nu \langle \partial_j\ov_i^\ell \partial_j\ov_i^\ell \rangle$ and $\eps_b^\ell = \eta \langle \partial_j\ob_i^\ell \partial_j\ob_i^\ell \rangle$ are the filtered kinetic and magnetic dissipation rates, respectively, and 
$
	\pi_D^\ell = - \left \langle(\partial_j\ob_i^\ell) \ov_i^\ell \ob_j^\ell \right \rangle = \left \langle(\partial_j\ov_i^\ell) \ob_i^\ell \ob_j^\ell \right \rangle = - \pi_M^\ell \ , 
$
are terms that  convert kinetic to magnetic energy $(\pi_D^\ell)$ and vice versa $(\pi_M^\ell)$, and 
\begin{align}
	\Pi_I^\ell &= - \left \langle(\partial_j\ov_i^\ell) \tij{I} \right \rangle \ , \label{piI} \\
	\Pi_M^\ell &= - \left \langle(\partial_j\ov_i^\ell) \tij{M} \right \rangle \ , \label{piM} \\
	\Pi_A^\ell &= - \left \langle(\partial_j\ob_i^\ell) \tij{A} \right \rangle \ , \label{piA}\\
	\Pi_D^\ell &= - \left \langle(\partial_j\ob_i^\ell) \tij{D} \right \rangle \label{piD}\ , 
\end{align}
denote the four  {\it proper}  energy fluxes, in the sense that they vanish in
the limit $\ell \to 0$, as can be seen from
eqs.~\eqref{eq:SGS_tensor_I}-\eqref{eq:SGS_tensor_D}.  If positive, the inertial and Maxwell
fluxes, $\Pi_I^\ell$ and $-\Pi_M^\ell$ transfer kinetic energy from scales
larger than or equal to $\ell$ to scales smaller than $\ell$ and vice versa if negative, while the
advective and dynamo fluxes, $\Pi_A^\ell$ and $-\Pi_D^\ell$, do so with
magnetic energy. Note that there is no interscale energy conversion as the
conversion terms $\pi_M^\ell$ and $\pi_D^\ell$ only involve filtered fields, as
such they are known as {\em resolved-scale conversion terms} \citep{Aluie17}.
The total energy flux is then given by the sum 
\be
\label{eq:flux_decomp}
    \Pi^\ell = \underbrace{-\pi_D^\ell - \pi_M^\ell}_{= \, 0 \ \forall \, \ell} + \Pi_I^\ell - \Pi_M^\ell + \Pi_A^\ell - \Pi_D^\ell \ .
\ee

\subsection{Flux decomposition: numerical results}
\label{sec:flux_dec_numerical}
\begin{figure*}[h]
    \centering
    \includegraphics[width = .98\columnwidth]{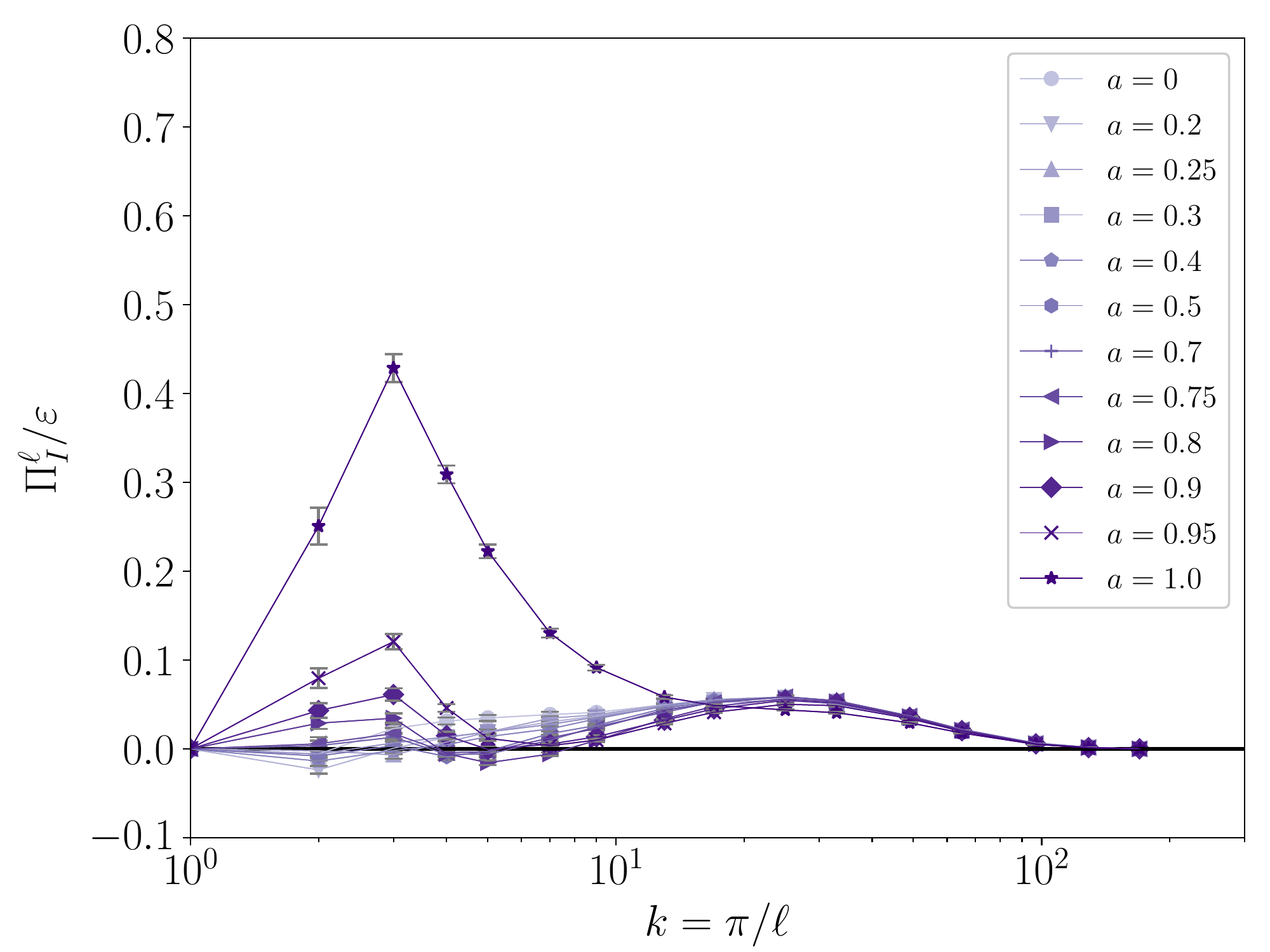}
	\includegraphics[width = .98\columnwidth]{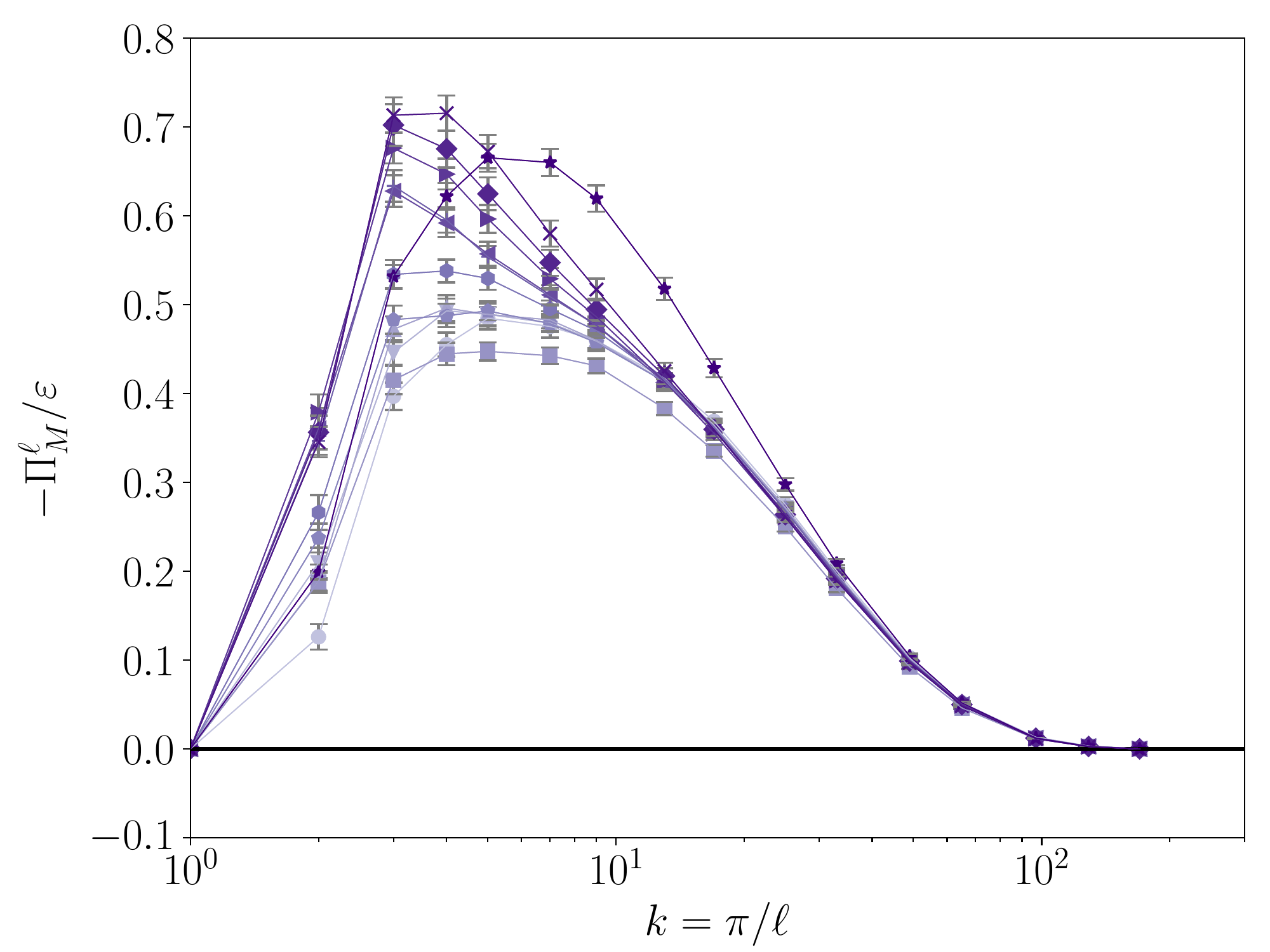}
	\includegraphics[width = .98\columnwidth]{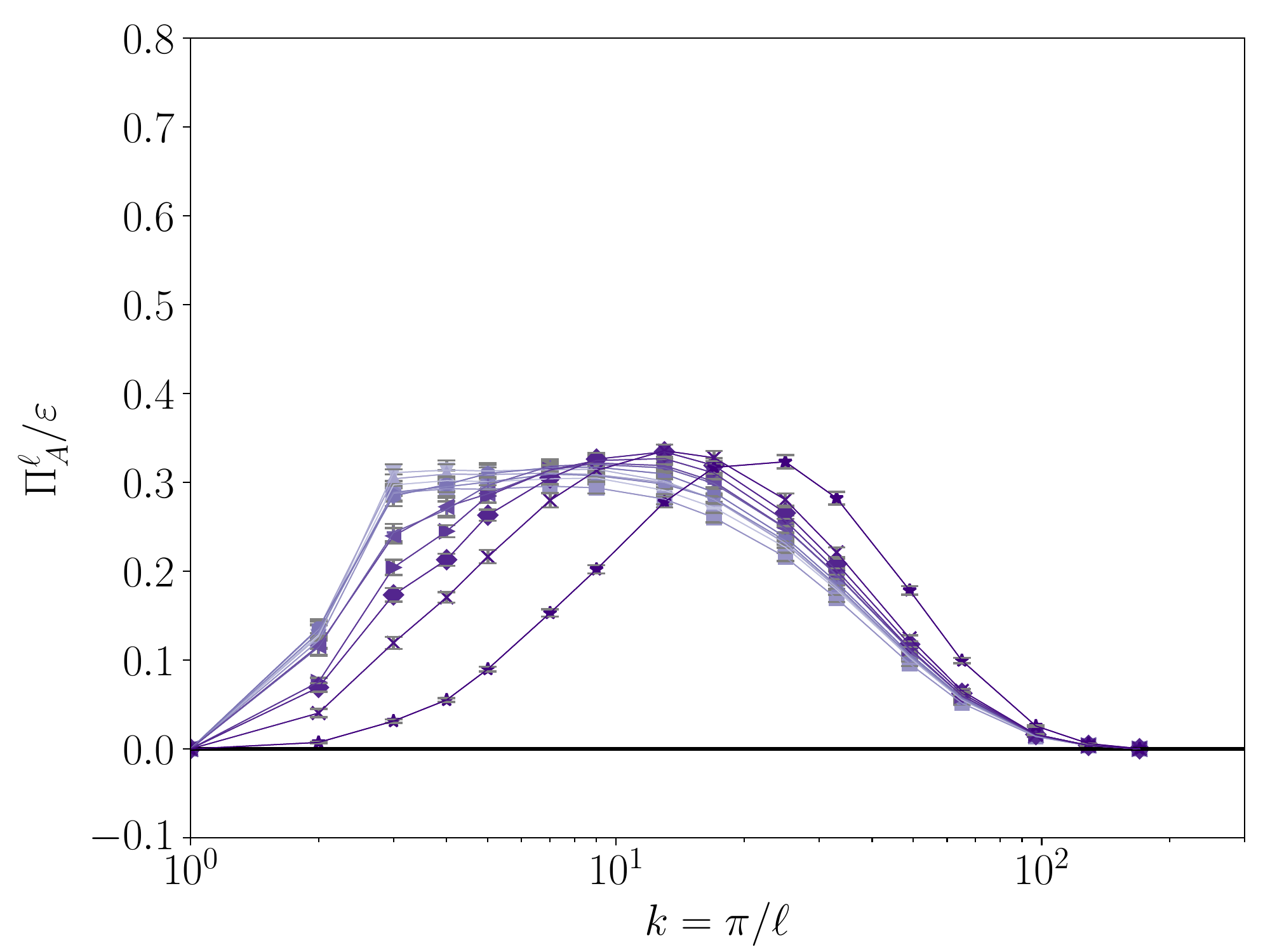}
	\includegraphics[width = .98\columnwidth]{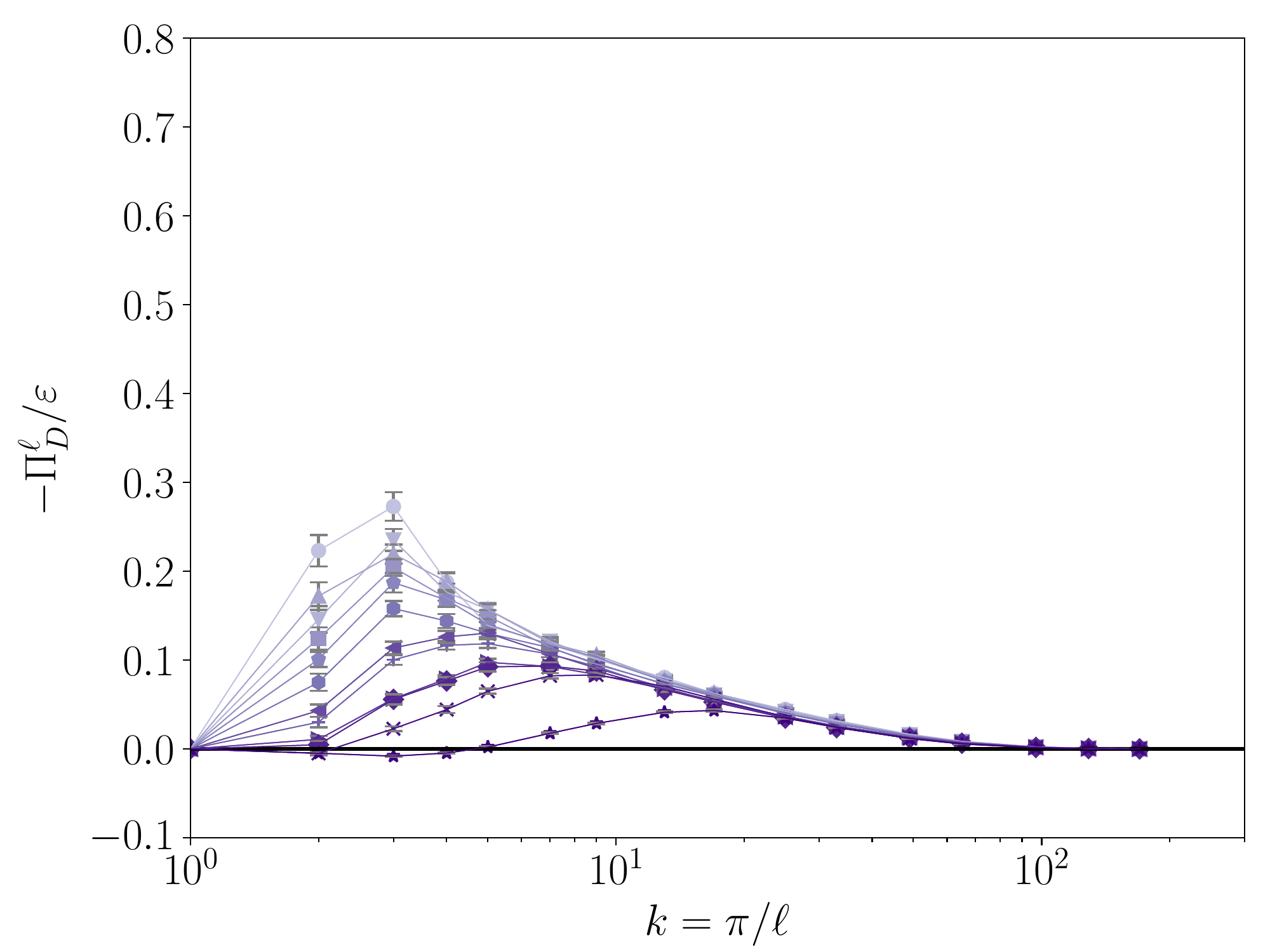}
	\caption{Decomposed fluxes for different values of $a$, normalized with the total energy dissipation rate $\eps=\eps_v+\eps_b$.
	Top left: $\Pi_I^\ell$, top right: $-\Pi_M^\ell$, 
	bottom left: $\Pi_A^\ell$, bottom right: $-\Pi_D^\ell$. 
	        The decreasing color gradient indicates decreasing values of $a$.
                The error bars show the standard error.}
	\label{fig:fluxes-new}
\end{figure*}

In this section we start by looking at the response of the four energy fluxes (\ref{piI}-\ref{piD}) at changing the forcing input parameter, $a$, as shown in  Fig. \ref{fig:fluxes-new}. 
As one can see, the first important result is given by the net transition of $\Pi_I^\ell$ when $a <1$, 
where the total energy transfer along this channel becomes almost vanishing at all scales (top left panel). 
A depletion of the inertial flux to almost zero in MHD turbulence subject to partially magnetic large-scale forcing 
had already been observed by \cite{alexakis2013large} at $Re_{\lambda}^{v} \approx 300$ for $a = 0.8$ and $a \approx 0.86$.
In both cases the magnetic forcing was helical, with a relative helicity $\rho = 0.31$ for $a = 0.8$ and $\rho = 0.5$ for 
$a \approx 0.86$. An injection of magnetic helicity and the subsequent local-in-scale generation of kinetic helicity \citep{Linkmann17triadnumerics} 
lead to a depletion of nonlinearity close to the injection scale. Here, both forcing functions are 
non-helical, hence the combination of the calculations by \cite{alexakis2013large} with those reported here  
suggest that the depletion of the inertial flux is a generic feature of magnetically forced MHD turbulence, irrespective 
of helicity injection.
As a result, the energy injected by the large-scale forcing on the velocity field is transferred to small scales only via the Lorentz contribution, i.e. the $-\Pi_M^\ell$ term (top right panel), which is pretty independent on the $a$ value.
The sign conventions for $\Pi_M^\ell$ and $\Pi_D^\ell$ reflect signs which occur in eqs.~\eqref{eq:evol_Eu} and \eqref{eq:evol_Eb}.
For $a \leqslant 0.5$ (magnetically dominated forcing) an inertial range in
$\Pi_A^\ell$ develops, indicating a direct cascade of magnetic fluctuations, see bottom left panel of Fig.~\ref{fig:fluxes-new}, typical of the physics of a passive vector advected by a turbulent flow and a signature that the direct contribution of the forcing on the induction equation introduces an important `linear' component in the magnetic transfer \citep{kraichnan1994anomalous,vergassola1996anomalous,chertkov1999small}.  At these
values of $a$ the inertial transfer $\Pi_I^\ell$ is very depleted despite the large-scale conversion of $b \to v$, see top left panel of Fig.~\ref{fig:fluxes-new}.
The  marked change  for $ a < 1$ have also a clear counterpart in the spectral 
scaling as seen by comparison of the energy spectra shown 
in Fig.~\ref{fig:energy_spectra} between $a=0.95$ and $a=1.00$.
The dynamo contribution $-\Pi_D^\ell$ (bottom right panel) does not develop any clear systematic plateau across wavenumbers. Its tendency to be more active at smaller and smaller $k$ by decreasing $a$ might results from the effect of increasing magnetic energy injection at large scales.

\section{Homochiral and heterochiral sub-channels}
\label{sec:homo-hetero}
\begin{figure*}[h]
    \centering
    \includegraphics[width = .98\columnwidth]{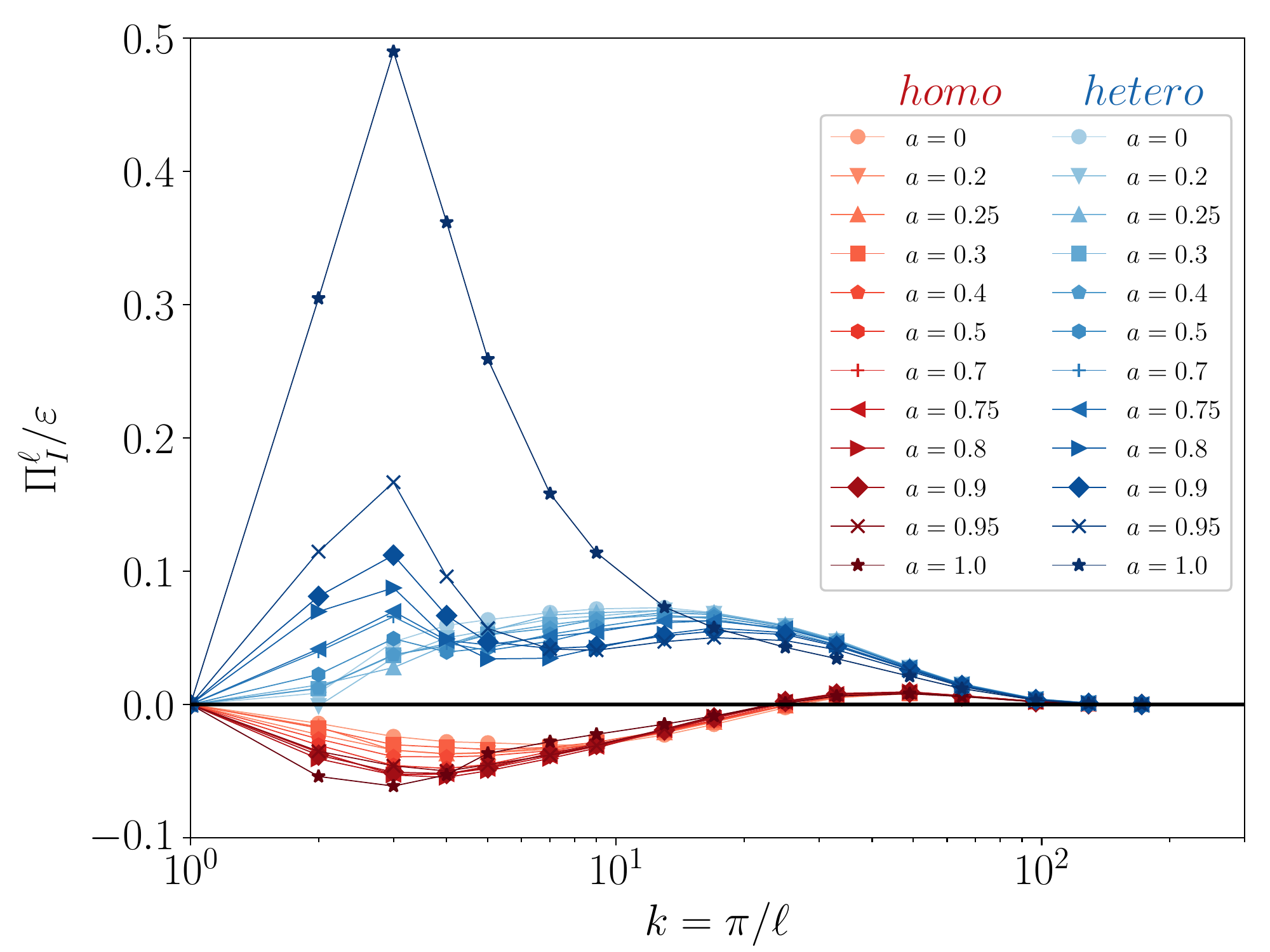}
	\includegraphics[width = .98\columnwidth]{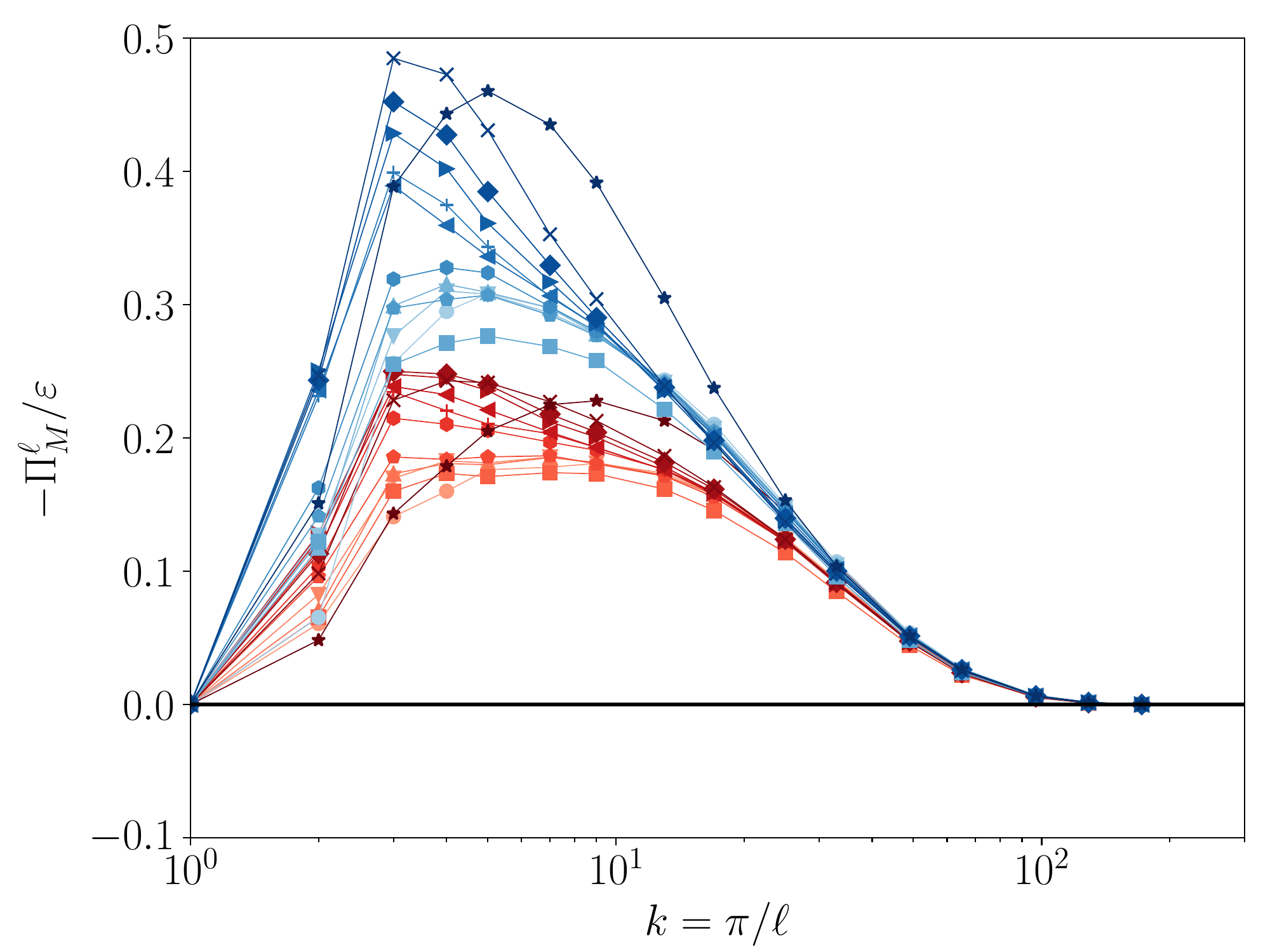}
	\includegraphics[width = .98\columnwidth]{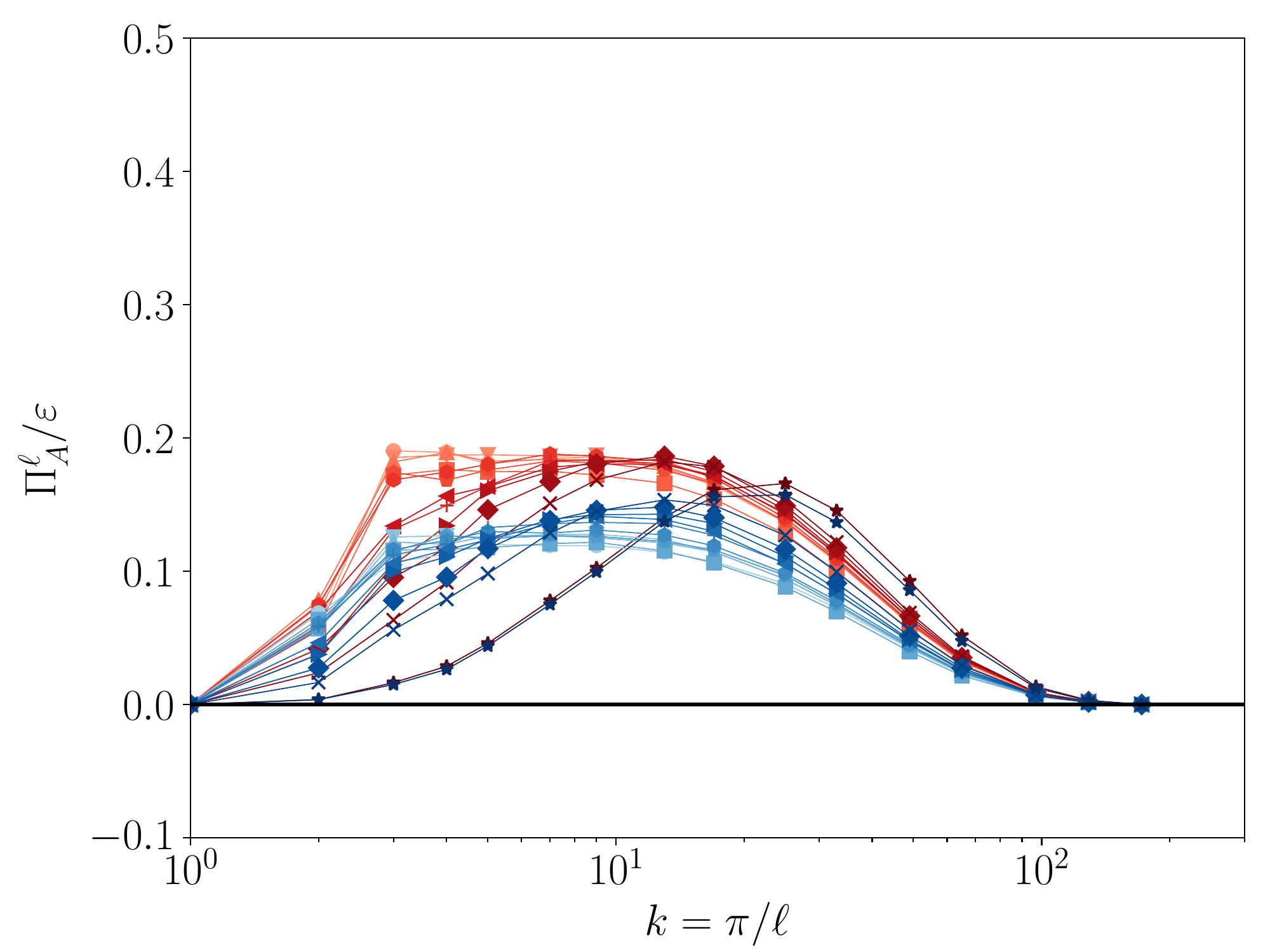}
	\includegraphics[width = .98\columnwidth]{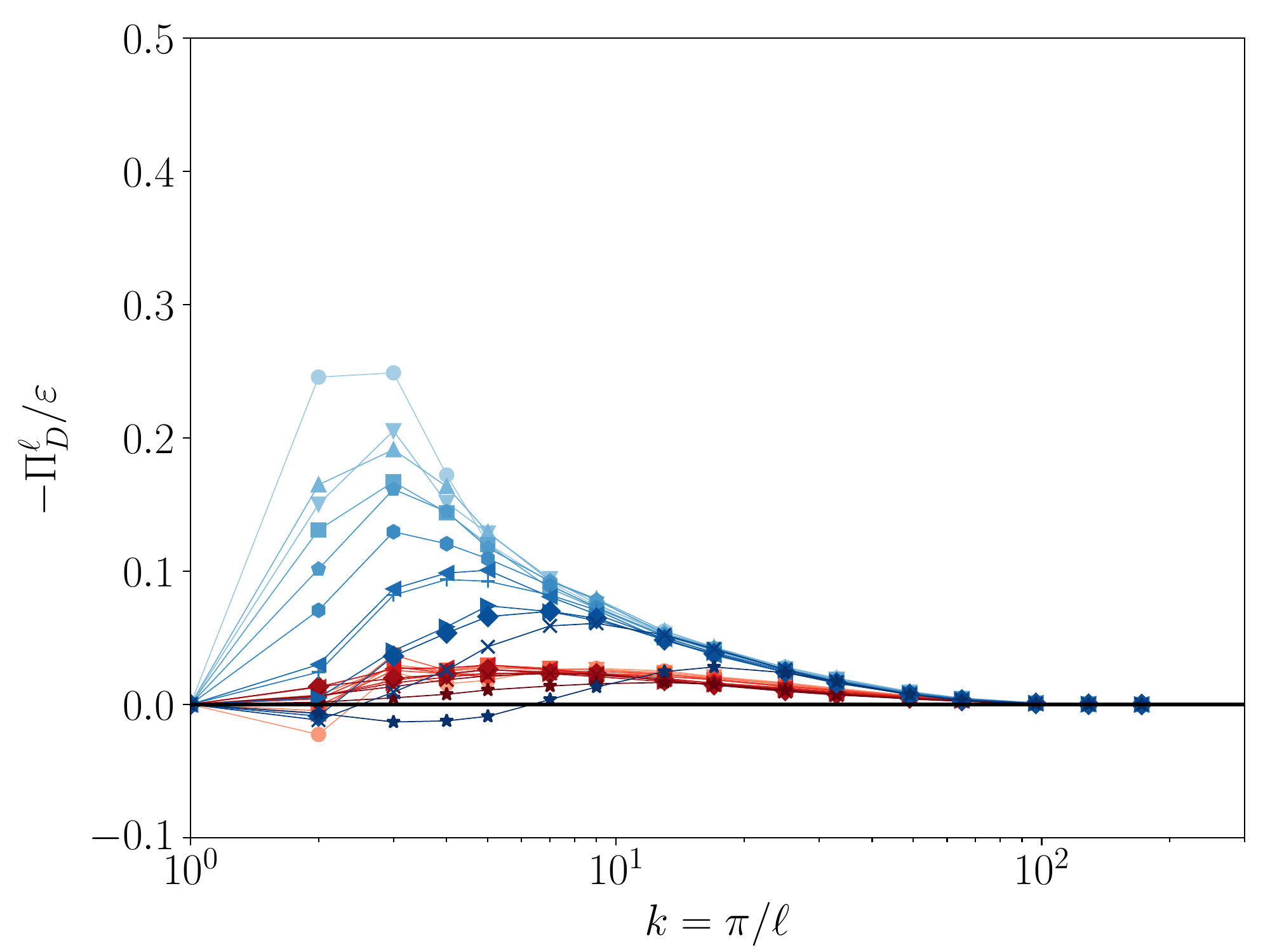}
	\caption{Hetero- and homochiral fluxes for different values of $a$.}
	\label{fig:pi-chiral}
\end{figure*}

Any solenoidal vector field can be decomposed in positively and negatively helical components \citep{waleffe92}:
\begin{align}
	\bm{v}(\bm{x},t) & = \bm{v}^+(\bm{x},t) + \bm{v}^-(\bm{x},t) \, \label{eq:helical_v}\\
	\bm{b}(\bm{x},t) & = \bm{b}^+(\bm{x},t) + \bm{b}^-(\bm{x},t) \ , \label{eq:helical_b}
\end{align}
where 
\begin{align}
\bm{v}^\pm(\bm{x},t) \equiv \sum_{\bm{k}} \left( \hat{\bm{v}}_{\bm{k}}(t) \cdot \bm{h}_{\bm{k}}^\pm \right) \bm{h}_{\bm{k}}^\pm  e^{i \bm{k} \cdot \bm{x} } \, \label{eq:def_helical_v} \\
\bm{b}^\pm(\bm{x},t) \equiv \sum_{\bm{k}} \left( \hat{\bm{b}}_{\bm{k}}(t) \cdot \bm{h}_{\bm{k}}^\pm \right) \bm{h}_{\bm{k}}^\pm  e^{i \bm{k} \cdot \bm{x} } \, \label{eq:def_helical_b} 
\end{align}
where $\bm{h}_{\bm{k}}^\pm$ 
are eigenfunctions of the curl operator $i\bm{k} \times (\cdot)$ with eigenvalues $\pm 1$. As a result, plugging the decomposition (\ref{eq:helical_v}-\ref{eq:helical_b}) in 
(\ref{eq:momentum_filtered}-\ref{eq:induction_filtered}) one gets different expressions for the spatio-temporally averaged energy sub-fluxes which can be reduced to homochiral or heterochiral contributions, depending whether the three fields entering in the expressions (\ref{piI}-\ref{piD}) have all the same chirality or there is one with opposite chirality of the other two (for details see Appendix A). As a result, the energy balance across scale can be further refined to:
\begin{align}
	\sqrt{a} \, \left \langle \ov_i^\ell \overline{{f_v}_i}^\ell \right  \rangle =&   \left (\Pi_I^{ho,\ell} + \Pi_I^{he,\ell} \right) - \left (\Pi_M^{ho,\ell} + \Pi_M^{he,\ell} \right)  \nonumber\\
	&-  \left (\pi_M^{ho,\ell} + \pi_M^{he,\ell} \right)  +   \eps_v^\ell\ , \label{eq:evol_Eu_hel}\\
	\sqrt{1-a} \, \left \langle \ob_i^\ell \overline{{f_b}_i}^\ell  \right \rangle  =&    \left (\Pi_A^{ho,\ell} + \Pi_A^{he,\ell} \right)  -   \left (\Pi_D^{ho,\ell} + \Pi_D^{he,\ell} \right) \nonumber \\
	&-  \left (\pi_D^{ho,\ell} + \pi_D^{he,\ell} \right)  +   \eps_b^\ell\ . \label{eq:evol_Eb_hel}
\end{align}
It is important to stress that all $\Pi$-labelled components are real fluxes, in the same sense of above, i.e.  they vanish for $\ell \to 0$.  
In Fig.~\ref{fig:pi-chiral} we show the helical decomposition of the four sub-fluxes shown  in Fig.~\ref{fig:fluxes-new}. By looking at all four panels of  Fig.~\ref{fig:pi-chiral}, we see that the heterochiral components tends to decrease for smaller and smaller $a$, except for the dynamo case $\Pi_D^\ell$.  Concerning the kinetic inertial channel, $\Pi_I^\ell$ (top left), it is important to observe that even in the presence of a vanishing global contribution (top left panel of Fig.~\ref{fig:fluxes-new}) the dynamic is far from being close to equilibrium, with a good  balancing between heterochiral contributions (forward flux) and homochiral ones (backward flux).  We are in the presence of a flux-loop cascade picture for this channel \citep{Alexakis2018cascades}. For the other two channels, $\Pi_M^\ell$ and $\Pi_A^\ell$ we do not observe any change in the flux direction with the heterochiral channel consistently more (less) important than the homochiral for  $\Pi_M^\ell$ ($\Pi_A^\ell$). Finally,  for $\Pi_D^\ell$ we observe that the  homochiral sector is  about 5\% of total $\Pi^{\ell}$, with a little variance with $a$ and the heterochiral is strongly depleted with increasing $a$. 

\subsection{Kinetic-magnetic conversion terms}
\label{sec:exchange}
\begin{figure}[h]
    \centering    
    \includegraphics[width = .98\columnwidth]{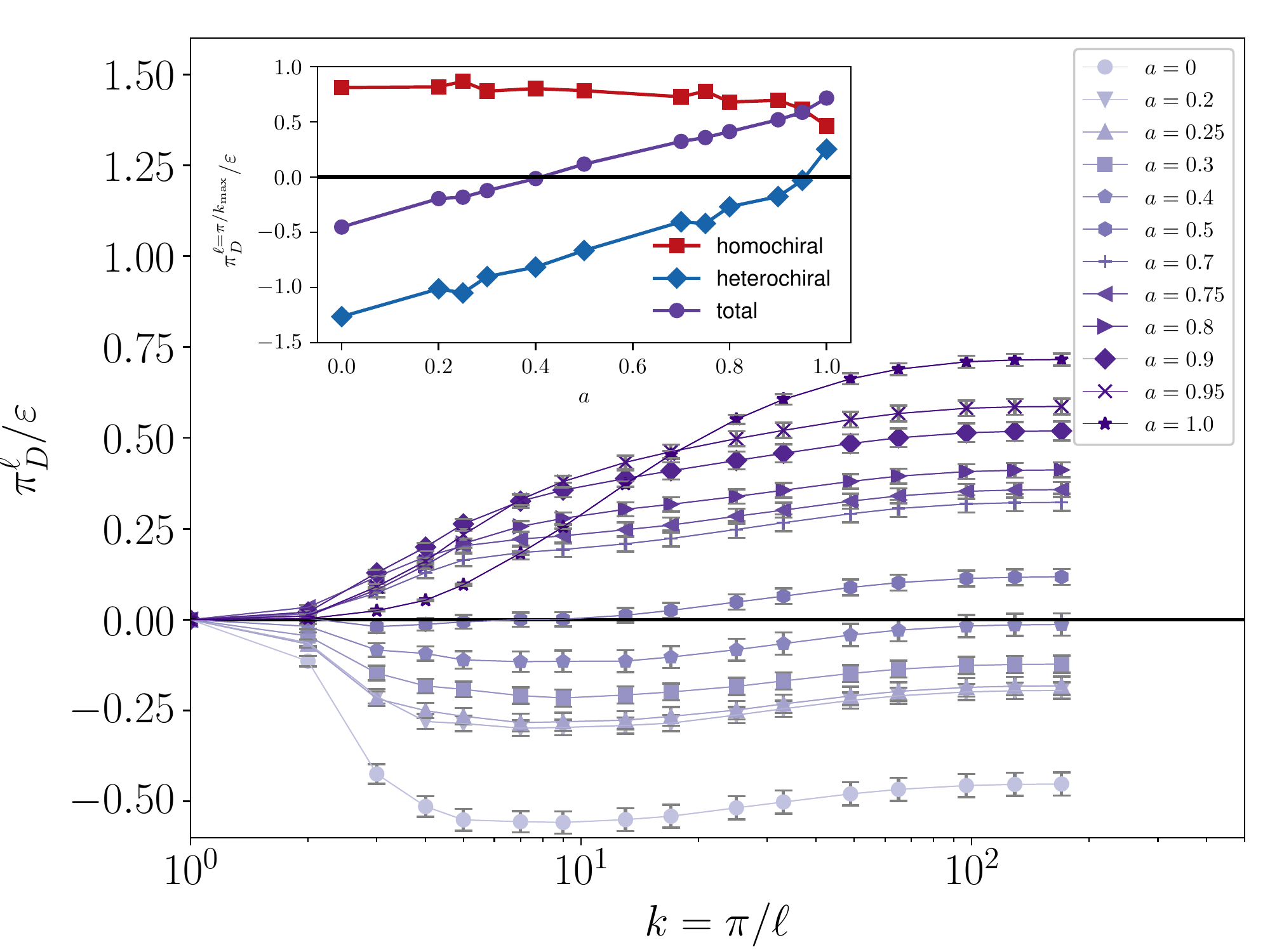}
	\includegraphics[width = .98\columnwidth]{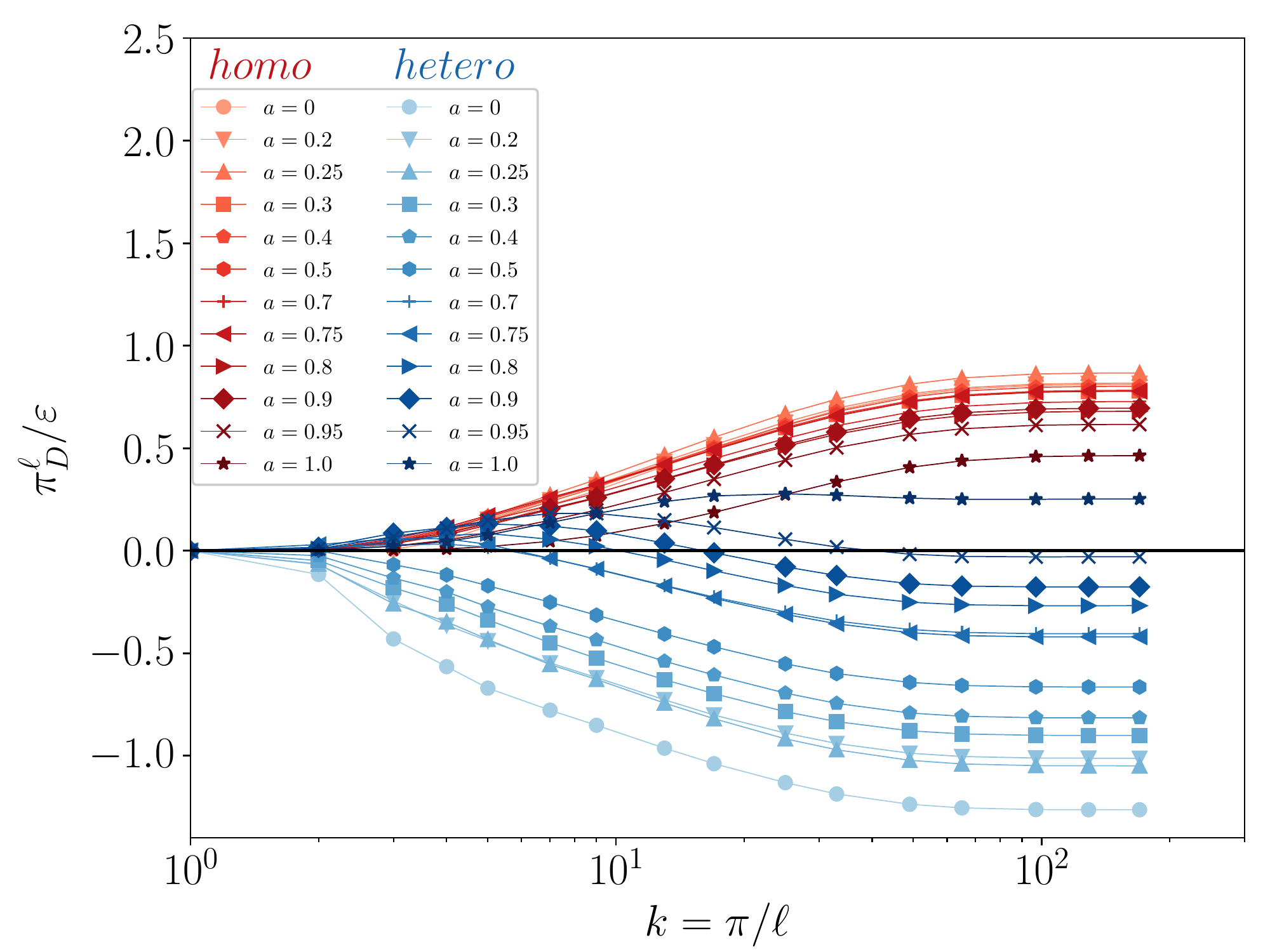}
	\caption{Top: $\pi_D^\ell$ versus the scale $k=\pi/\ell$. Bottom: decomposition of the curves shown in the top panel in hetero- and homochiral contributions as a function of wavenumber. Inset top panel: total transfer $\pi_D^\ell$ with $\ell=\pi/k_{\rm max}$ and its decomposition in hetero- and homochiral contributions.}
	\label{fig:conversion-chiral}
\end{figure}

\begin{figure}
	\centering
        \includegraphics[width = \columnwidth]{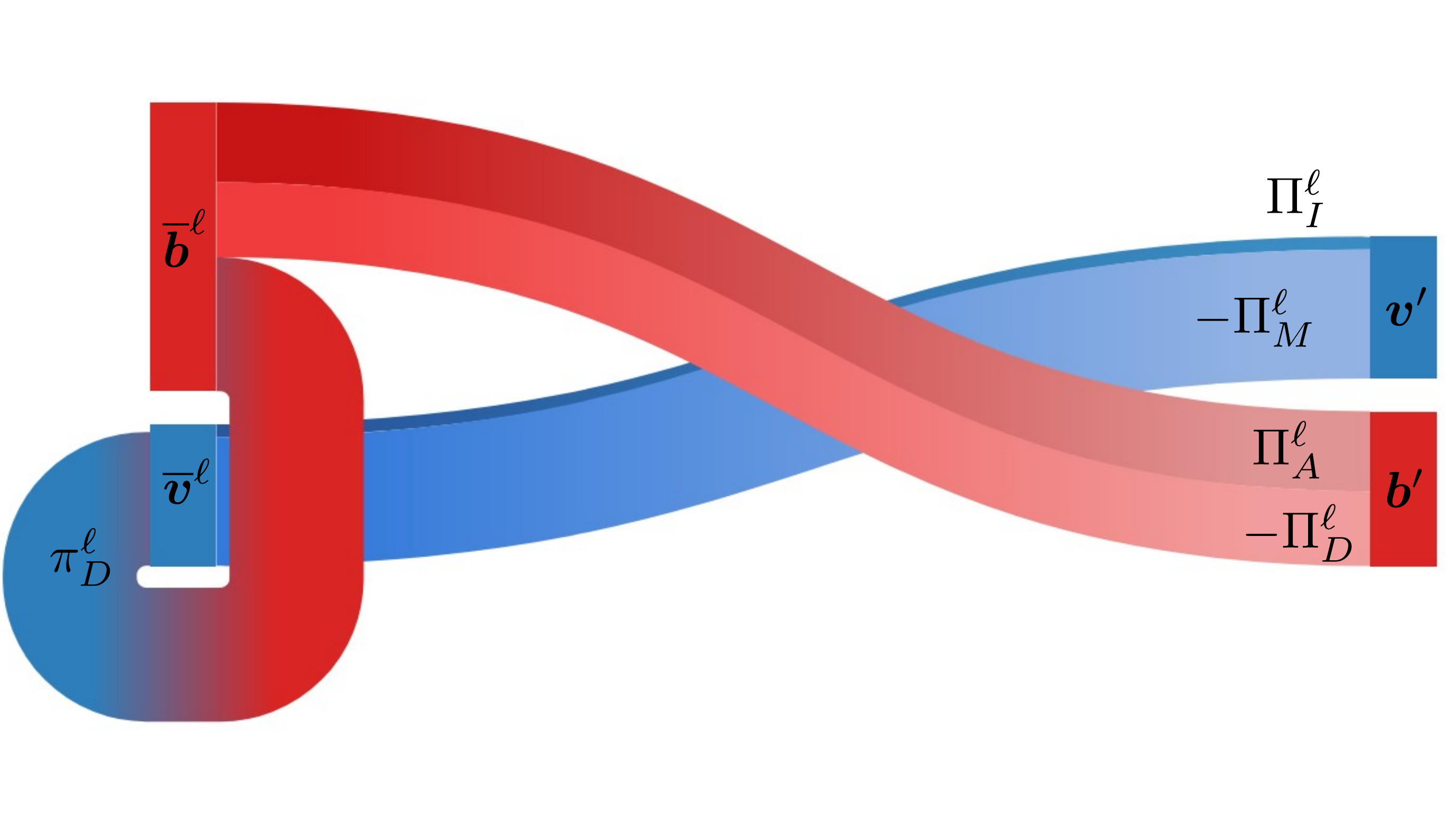}
	$a = 0$ 
        \includegraphics[width = \columnwidth]{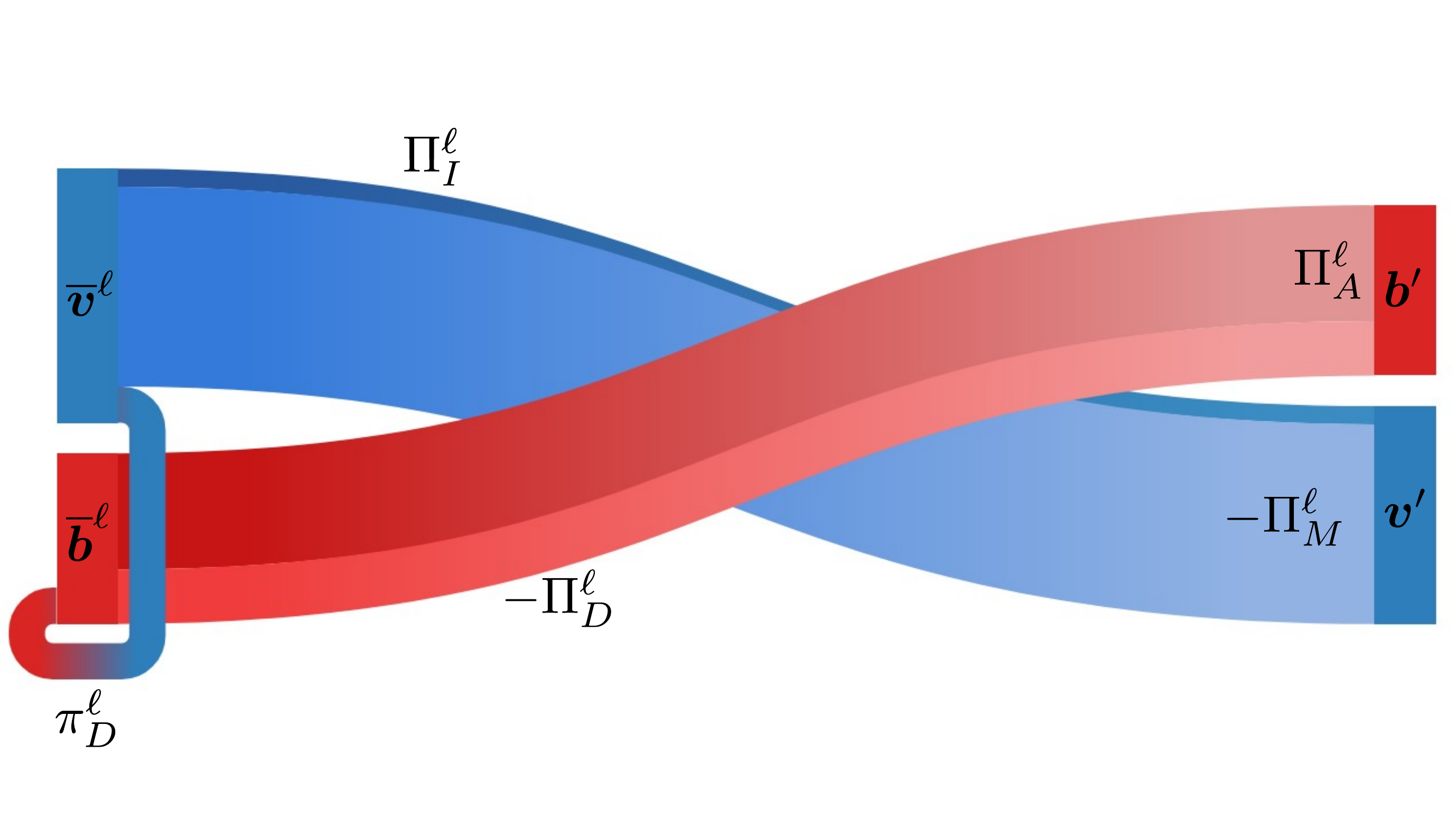}
	$a = 0.5$ 
        \includegraphics[width = \columnwidth]{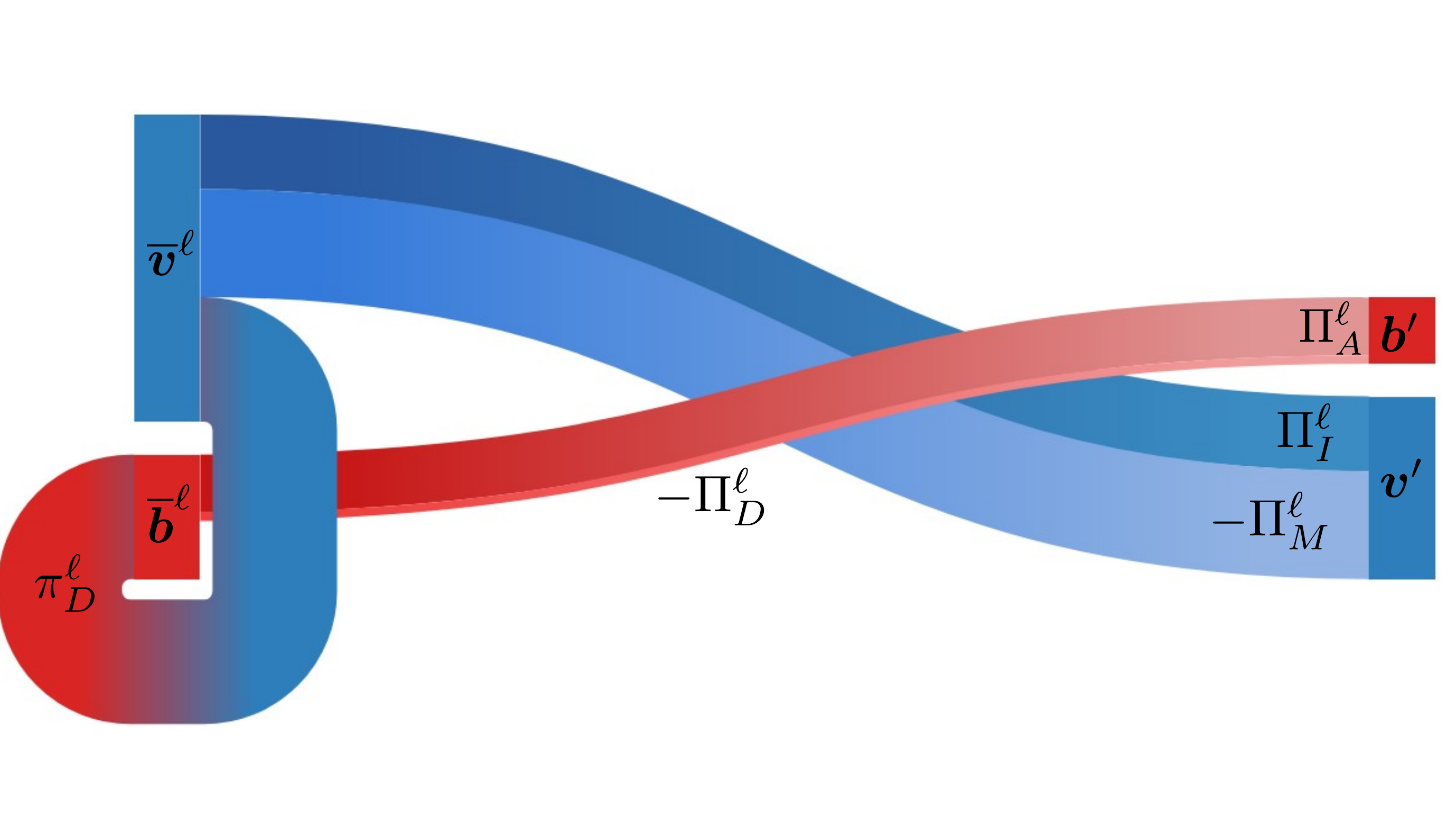} 
	$a = 1$
	\caption{
	Maximal energy fluxes and large-scale conversion terms for 
	$a = 0$, $a = 0.5$ and $a = 1$. The line widths correspond to 
	percentages of the total energy input $\eps$. The direction of the 
	transfers is indicated by the color gradient from dark to light. 
	Dark blue: $\max_\ell \Pi_I^\ell/\eps$,
	light blue: $\max_\ell -\Pi_M^\ell/\eps$,
	dark red: $\max_\ell \Pi_A^\ell/\eps$,
	light red: $\max_\ell -\Pi_D^\ell/\eps$,
	gradient blue to red: $\max_\ell |\pi_D^\ell|/\eps$. The sub-scale magnetic and 
	velocity field fluctuations are denoted by $\vec{b}^\prime = \vec{b} - \overline{\vec{b}}^\ell$ and 
	$\vec{v}^\prime = \vec{v} - \overline{\vec{v}}^\ell$, respectively.
	The diagrams have been created by adaptation of \citep{d3-circular}.
	}
	\label{fig:sankey}
\end{figure}

\begin{figure}
	\centering
        \includegraphics[width = \columnwidth]{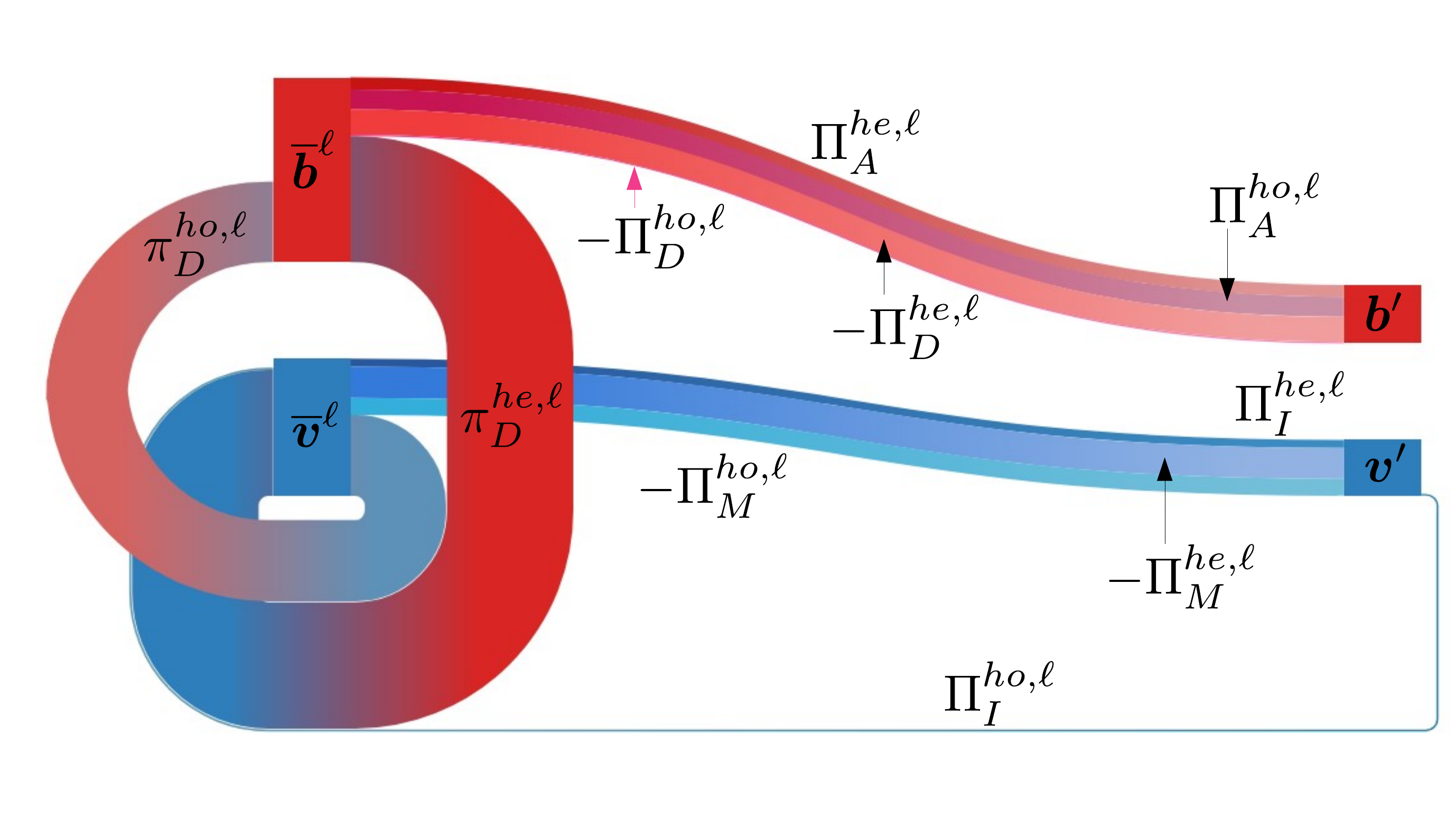} \\
	$a = 0$ \\ 
        \includegraphics[width = \columnwidth]{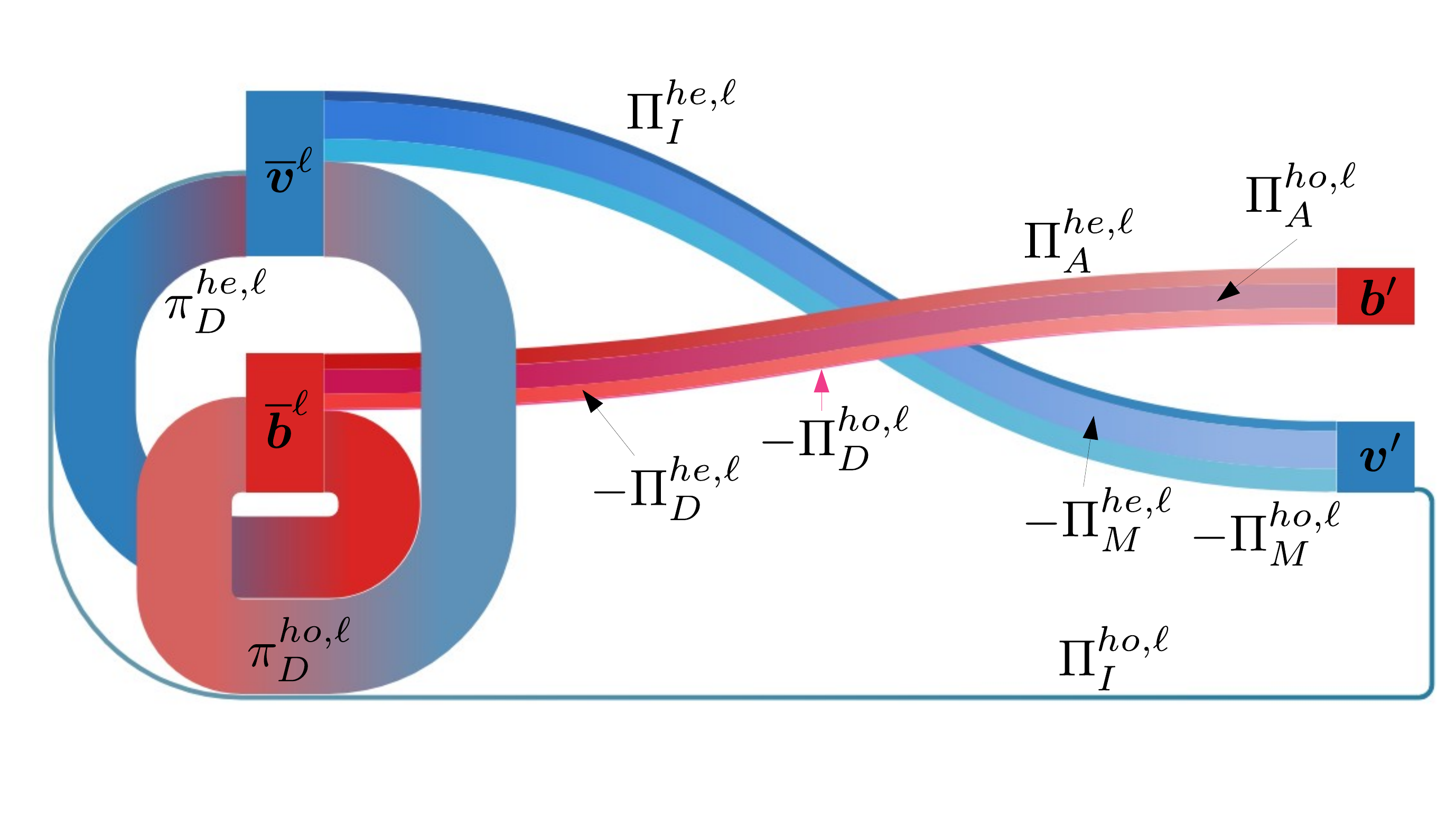} \\
	$a = 0.5$ \\
        \includegraphics[width = \columnwidth]{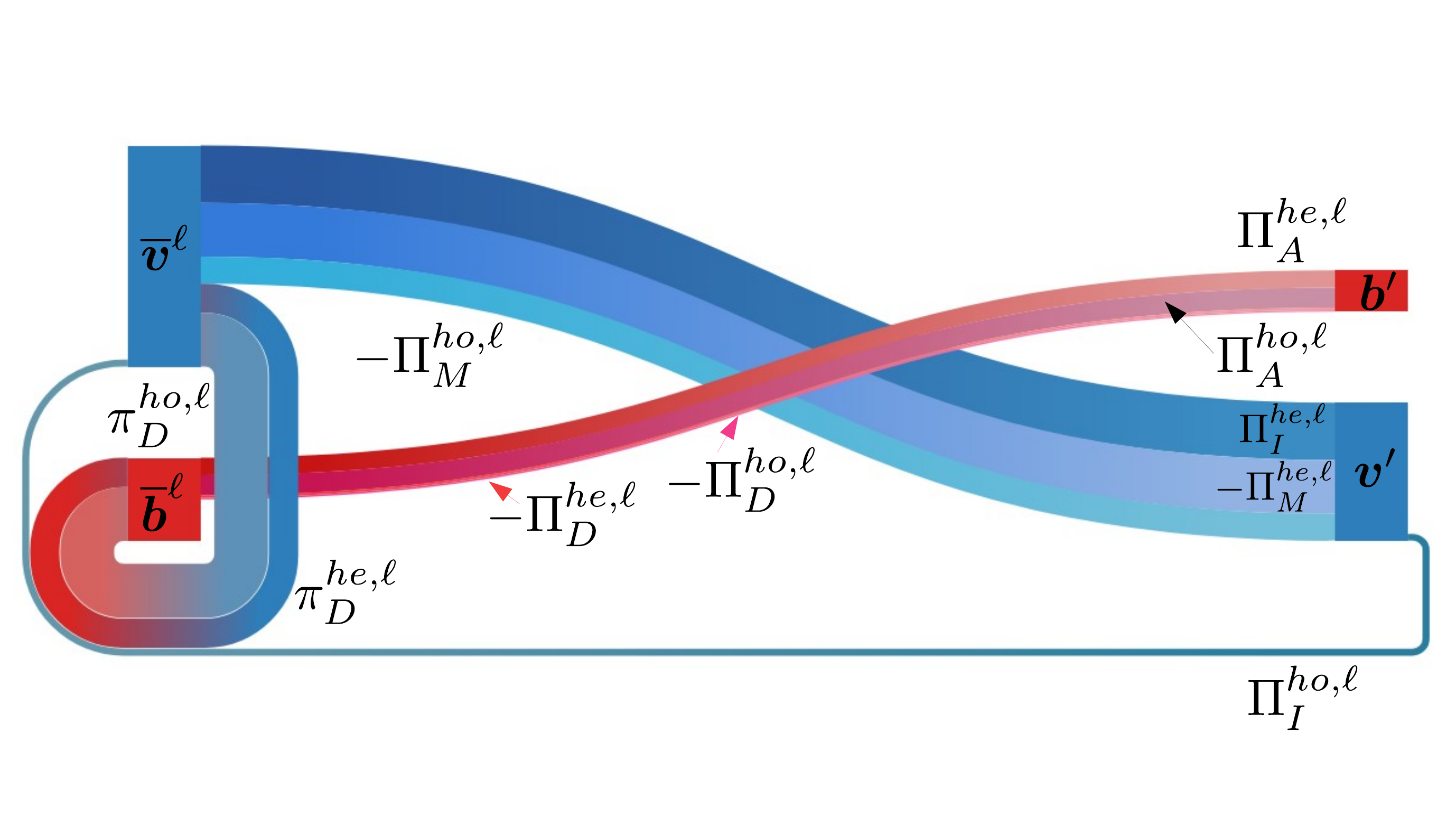} \\
	$a = 1$
	\caption{
	Helically decomposed maximal energy fluxes and large-scale conversion terms for 
	$a = 0$, $a = 0.5$ and $a = 1$. The line widths correspond to 
	percentages of the total energy input $\eps$.
	The direction of the
        transfers is indicated by the color gradient from dark to light.
	Dark blue: $\max_\ell \Pi_I^{he,\ell}/\eps$,
	dark cyan: $\max_\ell \Pi_I^{ho,\ell}/\eps$,
	light blue: $\max_\ell -\Pi_M^{he,\ell}/\eps$,
	light cyan: $\max_\ell -\Pi_M^{ho,\ell}/\eps$,
	dark red: $\max_\ell \Pi_A^{he,\ell}/\eps$,
	dark magenta: $\max_\ell \Pi_A^{ho,\ell}/\eps$,
	light red: $\max_\ell -\Pi_D^{he,\ell}/\eps$,
	light magenta: $\max_\ell -\Pi_D^{ho,\ell}/\eps$,
	gradient blue to red: $\max_\ell |\pi_D^{he,\ell}|/\eps$. 
	gradient light blue to light red: $\max_\ell |\pi_D^{ho,\ell}|/\eps$. 
	The sub-scale magnetic and
        velocity field fluctuations are denoted by $\vec{b}^\prime = \vec{b} - \overline{\vec{b}}^\ell$ and
        $\vec{v}^\prime = \vec{v} - \overline{\vec{v}}^\ell$, respectively.
	The diagrams have been created by adaptation of \citep{d3-circular}.
	}
	\label{fig:sankey-helical}
\end{figure}

The total conversion of kinetic to magnetic energy is quantified by the conversion term $\pi^\ell_D = -\pi^\ell_M$ in the limit $\ell \to 0$, in which eqs.~\eqref{eq:evol_Eu} and \eqref{eq:evol_Eb} become
\begin{align}
  \label{eq:evol_Eu_l0}
	  \pi_D^0  & = \sqrt{a} \, \left \langle \ov_i^0 \overline{{f_v}_i}^0 \right  \rangle  - \eps_v \ , \\
  \label{eq:evol_Eb_l0}
	- \pi_D^0  & =  \sqrt{1-a} \, \left \langle \ob_i^0 \overline{{f_b}_i}^0  \right \rangle  - \eps_b \ , 
\end{align}
as all terms in eqs.~\eqref{eq:evol_Eu} and \eqref{eq:evol_Eb} are continuous in $\ell$ and the flux terms vanish in the limit $\ell \to 0$.  
For the extreme cases, that is fully mechanical forcing, $a = 1$, and fully electromagnetic forcing, $a=0$, one obtains
\be
\pi^0_D = 
\begin{cases}
	\ \eps^{in}  - \eps_v > 0 & \ \text{for } a = 1 \ , \\
	- \eps^{in}  + \eps_b < 0 & \ \text{for } a = 0 \ , \\
\end{cases}
\ee
where $\eps^{in}$ is the spatio-temporally averaged total energy injection rate, which must equal the total dissipation during statistically stationary evolution, that is 
$\eps^{in} = \eps_v + \eps_b$. Hence, and as could have been expected, the energy conversion term will change sign as a function of $a$. If the forcing is fully mechanical, 
the magnetic field can only be maintained by dynamo action, that is, conversion of kinetic to magnetic energy encoded by $\pi_D^0 > 0$. For fully electromagnetic forcing, it is the other way round, 
the flow is maintained turbulent by the Lorentz force and $\pi_M^0 = -\pi_D^0 > 0$. 
In Fig.~\ref{fig:conversion-chiral} (top panel) we show the total undecomposed conversion term $\pi_D^\ell$ at changing $a$.
As one can see, while for mostly kinetic driving the conversion is finished at
mid-scale (as also shown by \cite{bian2019decoupled}), 
for magnetic forcing the conversion appears to be maximised at a  
much larger scale, as can be seen from the fact that  $\pi_D^\ell$ becomes almost constant for lower values of $k$, but in
total weaker than the $v \to b$ dynamo ($a = 1$).
As seen from the homochiral and heterochiral contributions shown in the bottom panel,  
on average the homochiral channel converts kinetic into magnetic energy and the
heterochiral one does the opposite.
Finally, in the inset of the top
panel we show the global conversion term, $\pi_D^\ell$ with $\ell=\pi/k_{max}$
as a function of $a$, showing that the main driving mechanisms upon changing the
forcing properties is linked to the heterochiral triads, which change quickly
as soon as $a <1$. The relative magnitudes 
of the different energy transfer and conversion terms for $a = 0$, $a = 0.5$ and $a = 1$ 
are visually summarised in form of Sankey diagrams in figs.~\ref{fig:sankey} and \ref{fig:sankey-helical}.
The former corresponds to the decomposition of the total energy flux as in 
eq.~\eqref{eq:flux_decomp} and the latter includes the helically decomposed terms as in 
eqs.~\eqref{eq:evol_Eu_hel} and \eqref{eq:evol_Eb_hel}.
As can be seen in Fig.~\ref{fig:sankey-helical}, the dynamo is always active, 
even for the case of purely magnetic forcing, as the homochiral interactions convert 
kinetic to magnetic energy for all values of $a$ through $\pi_D^{ho,\ell}$ shown in Fig.~\ref{fig:sankey-helical} 
in light purple, see also Fig.~\ref{fig:conversion-chiral}. The heterochiral conversion term  $\pi_D^{he,\ell}$ shown in Fig.~\ref{fig:sankey-helical} 
in dark purple, changes its role.
For $a = 1$, the magnetic field is sustained by the velocity field alone through both $\pi_D^{ho,\ell}$ and $\pi_D^{he,\ell}$, 
with decreasing $a$ the dynamo operates alongside the mean large-scale 
conversion of magnetic to kinetic energy due to the Lorentz force, which is associated with heterochiral
interactions, that is with $\pi_D^{he,\ell}$, for $a \neq 1$. The resulting 
{\em resolved-scale transfer loop} formed $\pi_D^{ho,\ell}$ and $\pi_D^{he,\ell}$ is clearly visible in fig.~\ref{fig:sankey-helical}
for $a=0$ and $a=0.5$. Interestingly, and in contrast to the dynamo for $a= 1$, 
the mean energy transfer due to the Lorentz force does vanish in both 
homo- and heterochiral transfer components  $\pi_D^{ho,\ell}$ and $\pi_D^{he,\ell}$ 
separately for $a = 0$.

The relation of homo- and heterochiral triads to energy transfer and
conversion in Navier-Stokes and MHD turbulence has been studied through shell models
\citep{Lessinnes09,DePietro2015,Rathmann2017,Rathmann2019mti}, 
analytical means \citep{waleffe92,Linkmann16,Linkmann17triadnumerics}, 
specifically designed DNS \citep{Biferale12,Linkmann17triadnumerics} and 
postprocessing of DNS data \citep{Alexakis17}. 
In particular in connection with kinematic
dynamo action, analytical results based on stability analyses of equilibria of
minimal representations of the MHD equations obtained by Galerkin truncation
suggest that homochiral triads do not contribute to the conversion of kinetic
to magnetic energy in the kinematic regime 
\citep{Linkmann16,Linkmann17triadnumerics}, both at large and small scale. 
Only heterochiral triads were found to contribute to the kinematic dynamo, which for 
a large-scale dynamo is commensurate with the helical signature of the 
$\alpha$-effect \citep{Steenbeck66,Brandenburg01}. 
Numerical results, obtained by 
projecting the velocity field on the positively helical subspace corroborate these 
findings, as the different helical sectors of the magnetic field mostly grow according to 
the helical signature of a stretch-twist-fold dynamo \citep{Linkmann17triadnumerics}.  
Similar results have recently been reported using specifically designed 
shell models of homo- or heterochiral triads and theoretical arguments based on the 
conservation of enstrophy-like invariants \citep{Rathmann2019mti}.
Interestingly, \cite{Rathmann2019mti} found that for homochiral triads the magnetic field could 
not be maintained by dynamo action alone, the magnetic dynamics had to be excited
by electromagnetic forcing, which confirms again that homochiral triads cannot 
amplify a magnetic seed field. 

In summary, in the kinematic regime \citet{Linkmann16,Linkmann17triadnumerics} and \cite{Rathmann2019mti}
report the {\em opposite} helical signature of that seen here for the fully nonlinear case ($a = 1$).  
In our nonlinear regime, the minimal homochiral Galerkin models do contain instabilities 
connected with dynamo action, which are not discussed in \citep{Linkmann16,Linkmann17triadnumerics}. 
The above discussion suggests that the process of dynamo saturation may be connected with a change in the behaviour of homo- and heterochiral triads. 

\subsection{Small-scale response}
\label{sec:small_scale}
\begin{figure}[h]
    \centering
	\includegraphics[width = .98\columnwidth]{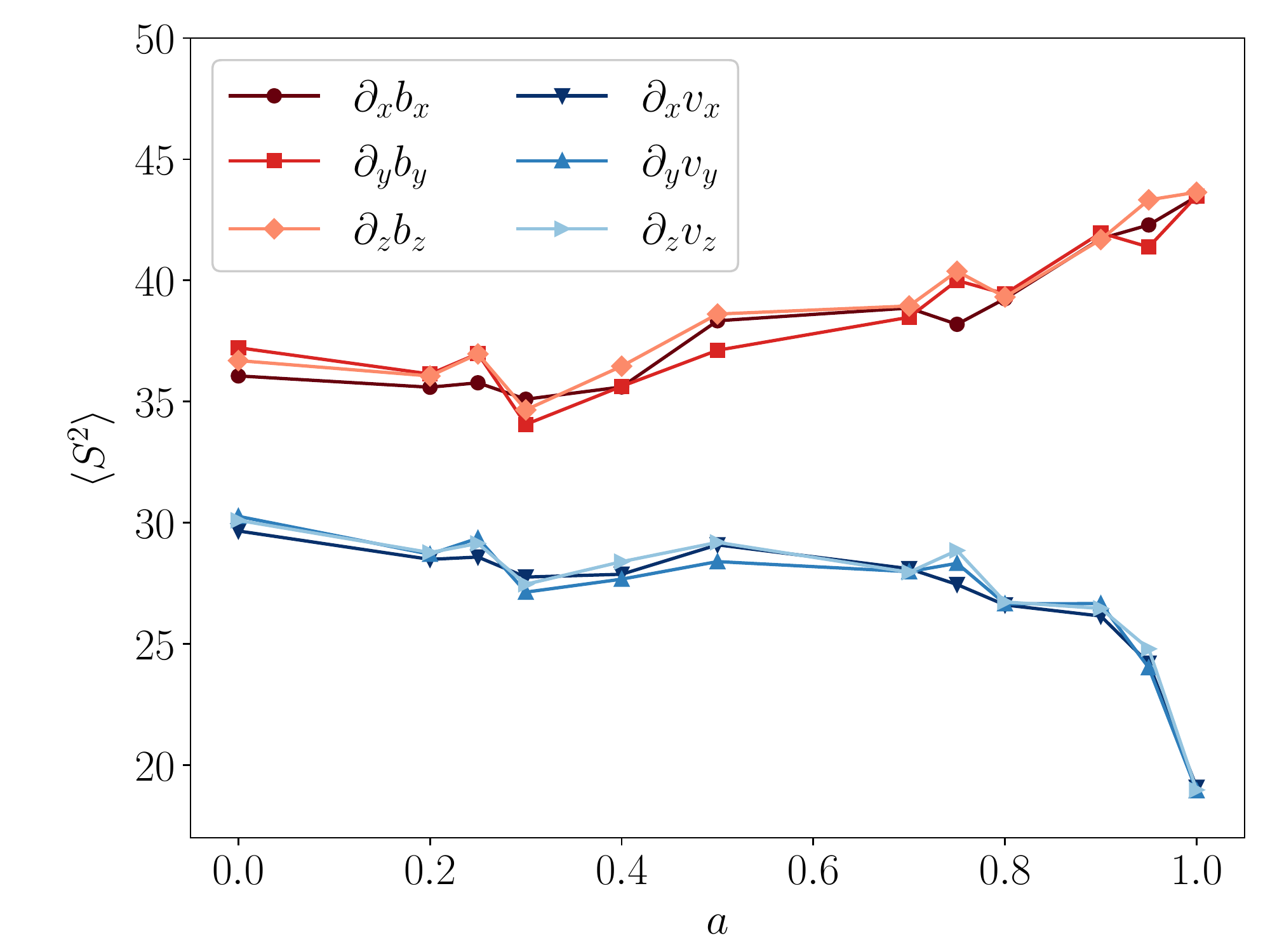}	
	\includegraphics[width = .98\columnwidth]{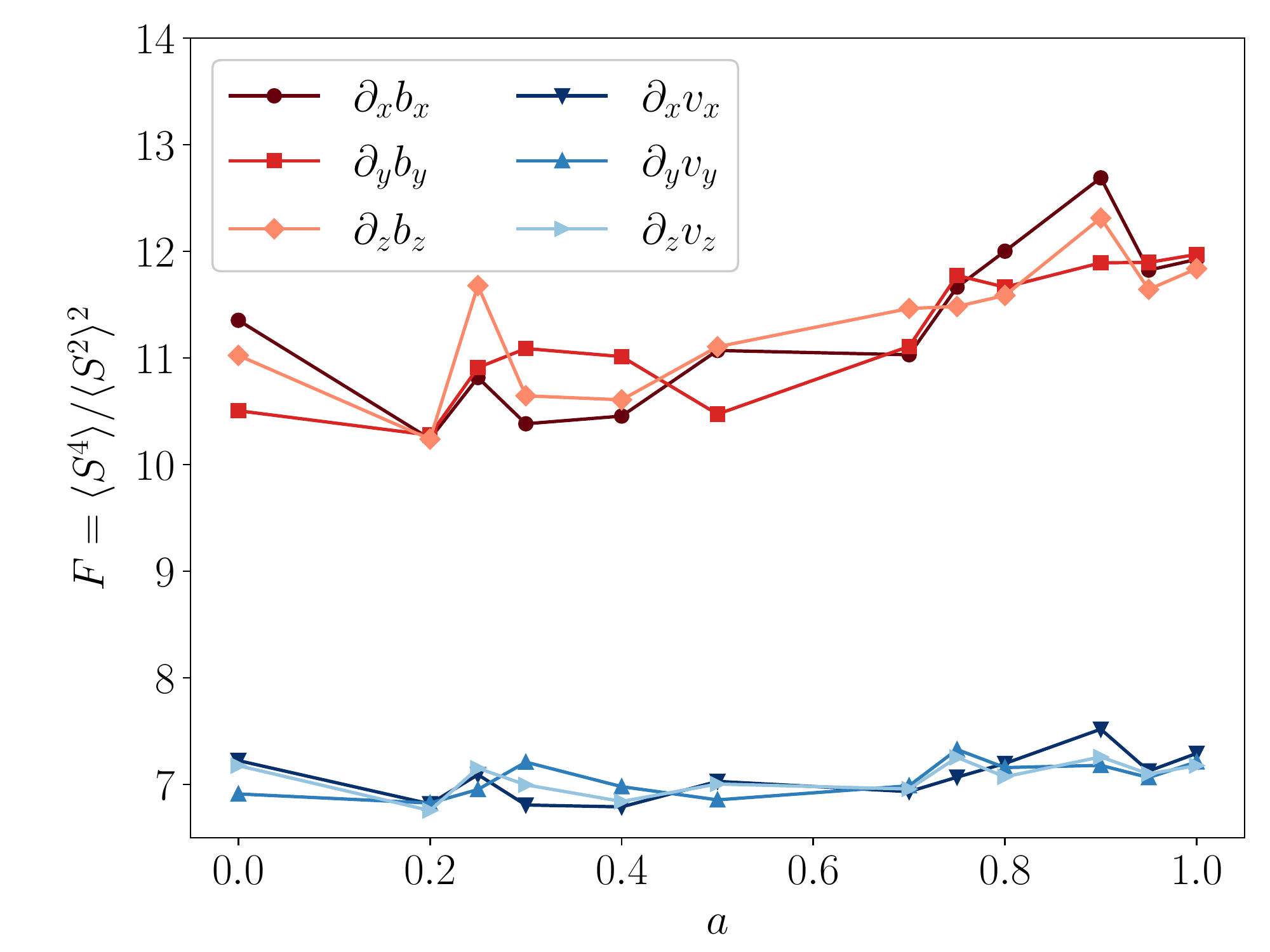}
	\caption{Top: second order moments of the three different longitudinal gradients for both velocity and magnetic field at changing $a$. Bottom: velocity and magnetic flatness for longitudinal gradients.}
	\label{fig:moments}
\end{figure}

In Fig. \ref{fig:moments} we summarise some results concerning the issue of universality of small-scales fluctuations. In the top panel we show the value of the second order moment for one component of the longitudinal gradient  for both velocity and magnetic fields, $\langle S^2_{v_i}\rangle=  \langle  (\partial_i v_i)^2 \rangle$ and  $\langle S^2_{b_i}\rangle= \langle  (\partial_i b_i)^2 \rangle$. As one can see, the main effect is given by an increase of the total magnitude of velocity gradients by decreasing $a$, indicating that the the depletion of the kinetic channel shown in Fig.~\ref{fig:fluxes-new} does not have a direct effect on the small-scale velocity activity. The bottom panel of the same figure shows that an even more universal behaviour is measured by looking at the flatness defined for the velocity field as:
$$
F_{v_i} = \frac{\langle (\partial_i v_i)^4 \rangle}{ \langle (\partial_i v_i)^2 \rangle^2}=\frac{\langle S^4_{v_i}\rangle}{\langle S^2_{v_i}\rangle^2}
$$
and similarly for the magnetic field. Both curves have a very small dependency on $a$, supporting the conclusion that small-scale fluctuations in MHD are strongly universal, at least concerning the kind of forcing variations studied in this paper.  
\section{Conclusions}
We have performed a systematic analysis of the total energy transfer in MHD at changing the large-scale forcing mechanisms, going from direct injection on the velocity field only to the
	case where stirring acts on the magnetic field only. We have split the total energy flux in 4 channels given by  (i) the kinetic  non-linear
	advection, (ii) the Lorentz force, (iii) the magnetic advection and
	(iv) magnetic stretching term and 2 sub-classes given by  heterochiral and homochiral components for a total of 8 different sub-fluxes. 
	We have shown that even a tiny injection of magnetic fluctuations at larger scales involves a quasi-singular response in the kinetic energy transfer mediated by the advection term, leading to an almost vanishing signal for the latter.  We also show that this negligible mean
	flux is the result  of a flux-loop balance between heterochiral (direct
	transfer) and homochiral (inverse transfer) channels. Conversely, both
	homochiral and heterochiral channels transfer energy forward for the
	other three channels. Furthermore,  by increasing the relative amount of
	magnetic injection we observe a reduction of the energy
	transferred via the heterochiral interactions for all channels with the 
	 exception of  the magnetic stretching channel.
	Cross exchange between velocity and magnetic field is reversed when the control forcing parameter is around. Small-scale properties are strongly universal, showing a non-trivial rearrangements of different transfer properties.  Our decomposition approach is exact and can be extended to study the flux of other relevant
	quantities, such as magnetic and cross helicities and it can be useful for
	improving sub-grid-modelling and to understand the energy transfer mechanisms in the presence of different injection mechanisms. 
\label{sec:conclusions}

\section*{Acknowledgements}
This work has been supported by NSFC grant Nos. 91752201, 11672123, and 11902138;
the Shenzhen Science and Technology Innovation Committee (grant No. KQTD20180411143441009) and the Department of Science and Technology of Guangdong Province (grant No. 2019B21203001).
We acknowledge computing support provided by the Center for Computational
Science and Engineering of Southern University of Science and Technology.
Calculations for this research were partly 
conducted on the Lichtenberg High-performance computer of the TU Darmstadt, Germany.
L.B. acknowledges the support from the Southern University of Science and Technology during
his visit when we started the research presented in this work.


\section{Appendix A: Helically decomposed fluxes and conversion terms}

As can be seen from eqs.~\eqref{eq:def_helical_v}-\eqref{eq:def_helical_b}, the
expansion of velocity and magnetic fields in their respective helical
components involves projections onto subspaces spanned by
$\bm{h}_{\bm{k}}^\pm$, 
and these operations clearly commute with the filtering
procedure defined in eq.~\eqref{eq:filtered_function}. As such, eq.~\eqref{eq:helical_v}
can be directly substituted into the filtered MHD equations to obtain filtered
evolution equations for the helical components. A subsequent projection onto
one helical subspace yields

\begin{align}
	 - \p{j} \sum_{s_2,s_3 \in \{+,-\}} \left(\overline{\ovsbi^\ell \ovscj^\ell}^\ell - \overline{\obsbi^\ell\obscj^\ell}^\ell 
			 + {\tij{I}}^{s_2s_3} - {\tij{M}}^{s_2s_3} \right)^{s_1} \nonumber\\
	 + \nu \p{jj} \ovsai^\ell + \sqrt{a} \, \overline{{f_v}_i^{s_1}}^\ell = \p{t}\ovsai^\ell \ , \label{eq:momentum_filtered_hel}\\
	 - \p{j} \sum_{s_2,s_3 \in \{+,-\}} \left(\overline{\obsbi^\ell \ovscj^\ell}^\ell  - \overline{\ovsbi^\ell \obscj^\ell }^\ell  
			 + {\tij{A}}^{s_2s_3} - {\tij{D}}^{s_2s_3}  \right)^{s_1} \nonumber\\
	 + \eta \p{jj} \obsai^\ell  +  \sqrt{1-a} \, \overline{{f_b}_i^{s_1}}^\ell = \p{t}\obsai^\ell \ , \label{eq:induction_filtered_hel}
\end{align}
where $s_1 \in \{+,-\}$ and we note the absence of a pressure term. Since the pressure gradient is orthogonal in the $L_2$-sense 
to the eigenfunctions $\bm{h}_{\bm{k}}^\pm$, the projection onto a helical subspace renders the nonlinear term solenoidal and removes the pressure gradient. The  
helically decomposed sub-filter-scale stress tensors read 
\begin{align}
	{\tij{I}}^{s_2s_3} =& \overline{v_i^{s_2} v_j^{s_3}}^\ell - \overline{\ovsbi^\ell \ovscj^\ell}^\ell \ , \label{eq:SGS_tensor_I_hel} \\
	{\tij{M}}^{s_2s_3} =& \overline{b_i^{s_2} b_j^{s_3}}^\ell - \overline{\obsbi^\ell \obscj^\ell}^\ell \ , \label{eq:SGS_tensor_M_hel} \\
	{\tij{A}}^{s_2s_3} =& \overline{b_i^{s_2} v_j^{s_3}}^\ell - \overline{\obsbi^\ell \ovscj^\ell}^\ell \ , \label{eq:SGS_tensor_A_hel} \\
	{\tij{D}}^{s_2s_3} =& \overline{v_i^{s_2} b_j^{s_3}}^\ell - \overline{\ovsbi^\ell \obscj^\ell}^\ell \ , \label{eq:SGS_tensor_D_hel} 
\end{align}
and following the same steps and in the derivation of the energy budget for the filtered components of velocity and magnetic field, 
one obtains four coupled energy budget equations for the filtered helical components 
\begin{align}
	0 = \frac{1}{2} \left \langle \frac{d}{dt} |\ovpi|^2 \right \rangle = 
								  & - \left (\Pi_I^{+++} + \Pi_I^{++-} + \Pi_I^{+-+} + \Pi_I^{+--} \right) \nonumber\\
								  & - \left (\pi_I^{+-+} + \pi_I^{+--} \right)  \nonumber \\ 
	                              & + \left ( \Pi_M^{+++} + \Pi_M^{++-} + \Pi_M^{+-+} + \Pi_M^{+--} \right) \nonumber\\
								  & + \left (\pi_M^{+++} + \pi_M^{++-} + \pi_M^{+-+} + \pi_M^{+--} \right)  \nonumber \\
								  & + \sqrt{a} \, \left \langle \ovpi^\ell \overline{{f_v}_i^+}^\ell \right  \rangle  \ , \label{eq:evol_Eup}\\
	0 = \frac{1}{2} \left \langle \frac{d}{dt} |\ovmi|^2 \right \rangle = 
	                               & - \left (\Pi_I^{---} + \Pi_I^{--+} + \Pi_I^{-+-} + \Pi_I^{-++} \right)  \nonumber\\
								   & - \left (\pi_I^{-+-} + \pi_I^{-++} \right)  \nonumber \\ 
	                               & + \left ( \Pi_M^{---} + \Pi_M^{--+} + \Pi_M^{-+-} + \Pi_M^{-++} \right)  \nonumber \\ 
								   & + \left (\pi_M^{---} + \pi_M^{--+} + \pi_M^{-+-} + \pi_M^{-++} \right)   \nonumber \\
								   & + \sqrt{a} \, \left \langle \ovmi^\ell \overline{{f_v}_i^-}^\ell \right  \rangle  \ , \label{eq:evol_Eum}\\
	0 = \frac{1}{2} \left \langle \frac{d}{dt} |\obpi|^2 \right \rangle = 
	                               & - \left (\Pi_A^{+++} + \Pi_A^{++-} + \Pi_A^{+-+} + \Pi_A^{+--} \right) \nonumber \\
								   & - \left (\pi_A^{+-+} + \pi_A^{+--} \right) \nonumber \\ 
	                               & + \left ( \Pi_D^{+++} + \Pi_D^{++-} + \Pi_D^{+-+} + \Pi_D^{+--} \right)  \nonumber \\
								   & + \left (\pi_D^{+++} + \pi_D^{++-} + \pi_D^{+-+} + \pi_D^{+--} \right)  \nonumber \\
								   & + \sqrt{1-a} \, \left \langle \obpi^\ell \overline{{f_b}_i^+}^\ell \right  \rangle  \ , \label{eq:evol_Ebp}\\
	0 = \frac{1}{2} \left \langle \frac{d}{dt} |\obmi|^2 \right \rangle = 
	                               & - \left (\Pi_A^{---} + \Pi_A^{--+} + \Pi_A^{-+-} + \Pi_A^{-++} \right) \nonumber \\
								   & - \left (\pi_A^{-+-} + \pi_A^{-++} \right)  \nonumber \\ 
	                               & + \left ( \Pi_D^{---} + \Pi_D^{--+} + \Pi_D^{-+-} + \Pi_D^{-++} \right)  \nonumber \\
								   & + \left (\pi_D^{---} + \pi_D^{--+} + \pi_D^{-+-} + \pi_D^{-++} \right)  \nonumber \\
								   & + \sqrt{1-a} \, \left \langle \obmi^\ell \overline{{f_b}_i^-}^\ell \right  \rangle  \, \label{eq:evol_Ebm}
\end{align}
where the superscript $\ell$ has been dropped to simplify the notation of the fluxes and conversion terms. The helically decomposed fluxes and resolved-scale conversion terms are defined as 
\begin{align}
	\Pi_I^{s_1 s_2 s_3} &= - \left \langle \left (\partial_j\ovsai^\ell \right ) {\tij{I}}^{s_2s_3} \right  \rangle \ , \label{eq:hel_PiI}\\
	\Pi_M^{s_1 s_2 s_3} &= - \left \langle \left (\partial_j\ovsai^\ell \right ) {\tij{M}}^{s_2s_3} \right \rangle \ , \label{eq:hel_PiM}\\
	\Pi_A^{s_1 s_2 s_3} &= - \left \langle \left (\partial_j\obsai^\ell \right ) {\tij{A}}^{s_2s_3} \right  \rangle \ , \label{eq:hel_PiA}\\
	\Pi_D^{s_1 s_2 s_3} &= - \left \langle \left (\partial_j\obsai^\ell \right ) {\tij{D}}^{s_2s_3} \right  \rangle \ , \label{eq:hel_PiD}
\end{align}
and 
\begin{align}
	\pi_I^{s_1 s_2 s_3} &= - \left \langle \left (\partial_j\ovsai^\ell \right ) \ovsbi^\ell \ovscj^\ell \right  \rangle \nonumber\\
	                    &= \left \langle \left (\partial_j\ovsbi^\ell \right ) \ovsai^\ell \ovscj^\ell \right  \rangle\nonumber\\
	                    &= - \pi_I^{s_2 s_1 s_3} \ , \label{eq:hel_conv_I}\\
	\pi_A^{s_1 s_2 s_3} &= - \left \langle \left (\partial_j\obsai^\ell \right ) \obsbi^\ell \ovscj^\ell \right  \rangle \nonumber\\
	                    &= \left \langle \left (\partial_j\obsbi^\ell \right ) \obsai^\ell \ovscj^\ell \right  \rangle \nonumber\\
	                    &= -\pi_A^{s_2 s_1 s_3}\ , \label{eq:hel_conv_A}\\
	\pi_D^{s_1 s_2 s_3} &= - \left \langle \left (\partial_j\obsai^\ell \right ) \ovsbi^\ell \obscj^\ell \right  \rangle \nonumber\\
	                    &= \left \langle \left (\partial_j\ovsbi^\ell \right ) \obsai^\ell \obscj^\ell \right \rangle \nonumber\\
	                    &= -\pi_M^{s_2s_1s_3}\ , \label{eq:hel_conv_MD}
\end{align}
A few observations can be made from these expressions and subsequently from eqs.~\eqref{eq:evol_Eup} - \eqref{eq:evol_Ebm}. First, eqs.~\eqref{eq:hel_conv_I} and \eqref{eq:hel_conv_A} imply that $\pi_I^{\pm\pm\pm}$, $\pi_I^{\pm\pm\mp}$,  
$\pi_A^{\pm\pm\pm}$, and  $\pi_A^{\pm\pm\mp}$ vanish, as these terms can be written as total gradients. Second, the remaining terms $\pi_I^{\pm\mp\pm} = -\pi_I^{\mp\pm\pm}$ exchange kinetic energy between $\overline{\bm{v}^+}$ and $\overline{\bm{v}^-}$, as 
can be seen from eqs.~\eqref{eq:evol_Eup} and \eqref{eq:evol_Eum}, which is the only kinetic energy exchange between positively and negatively helical sectors. The helically decomposed energy fluxes defined in eqs.~\eqref{eq:hel_PiI} - \eqref{eq:hel_PiD} conserve 
kinetic and magnetic energy for each helical component separately, which are categorized into homochiral or heterochiral fluxes, depending whether the three fields entering in the expressions \eqref{eq:hel_PiI} - \eqref{eq:hel_PiD} have all the same chirality or there is one with opposite chirality of the other two. It is trivial to obtain the energy balance equation for the filtered velocity and magnetic field by summing up eqs. ~\eqref{eq:evol_Eup} - \eqref{eq:evol_Ebm}, 
\begin{align}
	0 =& -  \left (\Pi_I^{ho,\ell} + \Pi_I^{he,\ell} \right) + \left (\Pi_M^{ho,\ell} + \Pi_M^{he,\ell} \right)  + \left (\pi_M^{ho,\ell} + \pi_M^{he,\ell} \right)  \nonumber\\
	   & + \sqrt{a} \, \left \langle \ov_i^\ell \overline{{f_v}_i}^\ell \right  \rangle  - \eps_v^\ell\ , \label{eq:evol_Eu_hel_app}\\
	0 =& -   \left (\Pi_A^{ho,\ell} + \Pi_A^{he,\ell} \right)  +   \left (\Pi_D^{ho,\ell} + \Pi_D^{he,\ell} \right)  +  \left (\pi_D^{ho,\ell} + \pi_D^{he,\ell} \right)  \nonumber\\
	   & + \sqrt{1-a} \, \left \langle \ob_i^\ell \overline{{f_b}_i}^\ell  \right \rangle  - \eps_b^\ell\ .  \label{eq:evol_Eb_hel_app}
\end{align}


\end{document}